\documentclass[a4paper,amsfonts,twocolumn,superscriptaddress,nofootinbib,longbibliography,accepted=2019-12-09]{quantumarticle}
\pdfoutput=1

\usepackage[numbers,sort&compress]{natbib}

\usepackage{epsfig,amsmath,amssymb,bm,epsf,graphicx,psfrag,bbm}
\usepackage{color, comment, verbatim}
\usepackage{ragged2e}
\usepackage{subfigure}
\usepackage{enumerate}
\usepackage{chngcntr}
\usepackage[colorlinks]{hyperref}
\usepackage[figure,table]{hypcap}

\definecolor{orange}{RGB}{255,153,51}

\hypersetup{
	bookmarksnumbered,
	pdfstartview={FitH},
	citecolor={darkgreen},
	linkcolor={darkred},
	urlcolor={darkblue},
	pdfpagemode={UseOutlines}}
\definecolor{darkgreen}{RGB}{50,190,50}
\definecolor{darkblue}{RGB}{0,0,190}
\definecolor{darkred}{RGB}{238,0,0}
\usepackage{soul}

\definecolor{fuchsia}{rgb}{1.0, 0.0, 1.0}
\definecolor{orange}{rgb}{1.0, 0.4, 0.0}

\def\ket#1{\vert#1\rangle}
\usepackage{amsthm}

\usepackage[framemethod=tikz]{mdframed}
\definecolor{mycolor}{RGB}{83,37,127}%{0.122, 0.435, 0.698}
\newmdenv[innerlinewidth=0.5pt, roundcorner=4pt,linecolor=mycolor,innerleftmargin=6pt,
innerrightmargin=6pt,innertopmargin=6pt,innerbottommargin=6pt]{mybox}
\usepackage{paracol}
\usepackage{tcolorbox}
\tcbset{colback=mycolor!5,colframe=mycolor,
fonttitle=\bfseries, float=htb}
\newtcolorbox{smallbox}[3][]
{float=ht!,lower separated=false, blend before title=colon hang,
title={#2}, label= fig:#3 ,#1}
\usepackage{enumitem}

\makeatletter
\def\blfootnote{\xdef\@thefnmark{}\@footnotetext}
\makeatother

\begin{document}

\title{Optimizing Quantum Error Correction Codes with Reinforcement Learning}
%%%
\author{Hendrik Poulsen Nautrup}
\email{hendrik.poulsen-nautrup@uibk.ac.at}
\affiliation{Institute for Theoretical Physics, University of Innsbruck, Technikerstr. 21a, A-6020 Innsbruck, Austria}
\orcid{0000-0001-7815-7006}
\author{Nicolas Delfosse}
%\email{}
\affiliation{Station Q Quantum Architectures and Computation Group, Microsoft Research, Redmond, WA 98052, USA}
\orcid{0000-0002-3949-981X}
\author{Vedran Dunjko}
%\email{}
\affiliation{LIACS, Leiden University, Niels Bohrweg 1, 2333 CA Leiden, The Netherlands}
\orcid{0000-0002-2632-7955}
\author{Hans J. Briegel}
%\email{hans.briegel@uibk.ac.at}
\affiliation{Institute for Theoretical Physics, University of Innsbruck, Technikerstr. 21a, A-6020 Innsbruck, Austria}
\affiliation{Department of Philosophy, University of Konstanz, Konstanz 78457, Germany}
\orcid{0000-0002-9065-1565}
\author{Nicolai Friis}
%\email{nicolai.friis@univie.ac.at}
\affiliation{Institute for Quantum Optics and Quantum Information, Austrian Academy of Sciences, Boltzmanngasse 3, 1090 Vienna, Austria}
\affiliation{Institute for Theoretical Physics, University of Innsbruck, Technikerstr. 21a, A-6020 Innsbruck, Austria}
\orcid{0000-0003-1950-8640}
%%
%%%%
\date{December 13, 2019}
\begin{abstract}
Quantum error correction is widely thought to be the key to fault-tolerant quantum computation.
However, determining the most suited encoding for unknown error channels or specific laboratory setups is highly challenging. Here, we present a reinforcement learning framework for optimizing and fault-tolerantly adapting quantum error correction codes.
We consider a reinforcement learning agent tasked with modifying a family of surface code quantum memories until a desired logical error rate is reached. Using efficient simulations with about 70 data qubits with arbitrary connectivity, we demonstrate that such a reinforcement learning agent can determine near-optimal solutions, in terms of the number of data qubits, for various error models of interest. Moreover, we show that agents trained on one setting are able to successfully transfer their experience to different settings. This ability for transfer learning showcases the inherent strengths of reinforcement learning and the applicability of our approach for optimization from off-line simulations to on-line laboratory settings.
\end{abstract}
\maketitle
\section{Introduction}
\vspace*{-2mm}

Quantum computers hold the promise to provide advantages over their classical counterparts for certain classes of problems~\cite{NielsenChuang2000, DunjkoGeCirac2018, CampbellKhuranaMontanaro2019}. Yet, such advantages may be fragile and can quickly disappear in the presence of noise, losses, and decoherence. Provided the noise is below a certain threshold, these difficulties can in principle be overcome by means of fault-tolerant quantum computation~\cite{NielsenChuang2000, Preskill1997}. There, the approach to protect quantum information from detrimental effects is to encode each logical qubit into a number of data qubits. This is done in such a way that physical-level errors can be detected, and corrected, without affecting the logical level, provided they are sufficiently infrequent~\cite{GottesmanPhD1997}.
Quantum error correction (QEC) codes thus allow for devices \textemdash usually referred to as quantum memories~\cite{Terhal2015} \textemdash that can potentially store quantum information for arbitrarily long times if sufficiently many physical qubits are available. However, physical qubits will be a scarce resource in near-term quantum devices. It is hence desirable to make use of QEC codes that are resource-efficient given a targeted logical error rate.
Yet, while some types of errors can be straightforwardly identified and corrected, determining the most suitable QEC strategy for arbitrary noise is a complicated optimization problem. Nevertheless, solutions to this complex problem may offer significant advantages, not only in terms of resource efficiency but also error thresholds~\cite{TuckettBartlettFlammia2018, FujiiTokunaga2012}.

Here, we consider a scenario where certain QEC codes can be implemented on a quantum memory that is subject to arbitrary noise. Given the capacity to estimate the logical error rate, our objective is to provide an automated scheme that determines the most economical code that achieves a rate below a desired threshold. A key contributing factor to the complexity of this task is the diversity of the encountered environmental noise and the corresponding error models. That is, noise may not be independent and identically distributed, may be highly correlated or even utterly unknown in specific realistic settings~\cite{Monz-Blatt2011, Schindler-Blatt2013}. Besides possible correlated errors, the error model might change over time, or some qubits in the architecture might be more prone to errors than others. Moreover, even for a given noise model, the optimal choice of QEC code still depends on many parameters such as the minimum distance, the targeted block error rate, or the computational cost of the decoder~\cite{NielsenChuang2000}. Determining these parameters requires considerable computational resources. %, e.g., the computation of the minimum distance of a linear code is NP-hard~\cite{Vardy1997}.
At the same time, nascent quantum computing devices are extremely sensitive to noise while having only very few qubits available to correct errors.

For the problem of finding optimized QEC strategies for near-term quantum devices, adaptive machine learning~\cite{DunjkoBriegel2018} approaches may succeed where brute force searches fail. In fact, machine learning has already been applied to a wide range of decoding problems in QEC~\cite{TorlaiMelko2017, KrastanovJiang2017, VarsamopoulosCrigerBertels2017, BaireutherOBrienTarasinskiBeenakker2018, BreuckmannNi2018, ChamberlandRonagh2018, SwekeKesselringVanNieuwenburgEisert2018, BaireutherCaioCrigerBeenakkerOBrien2018, Ni2018, MaskaraKubicaOConnor2018, LiuPoulin2018, DavaasurenSuzukiFujiiKoashi2018, AndreassonJohanssonLiljestrandGranath2019, VarsamopoulosBertelsAlmudever2019a, VarsamopoulosBertelsAlmudever2019b, DomingoColomerSkotiniotisMunozTapia2019, WagnerKampermannBruss2019, ChinniKulkarniPaiMitraSarvepalli2019, ShethJafarzadehGheorghiu2019}. Efficient decoding is of central interest in any fault-tolerant scheme. However, only a limited improvement can be obtained by optimizing the decoding procedure alone since decoders are fundamentally limited by the underlying code structure. Moreover, a number of efficient decoders already exist for topological codes~\cite{DelfosseIyerPoulin2016, SQUABweb, DelfosseNickerson2017}. This paper deals with an entirely different question: Instead of considering modifications of an algorithm run on a classical computer to support a quantum computation, we ask how the structure of the quantum mechanical system itself can be changed, both a priori and during a computation. More specifically, we present a reinforcement learning (RL)~\cite{SuttonBarto1998} framework for adapting and optimizing QEC codes which we apply to a family of surface codes in a series of simulations. Through an adaptive code selection procedure, our approach enables tailoring QEC codes to realistic noise models for state-of-the-art quantum devices and beyond.
%
%However, all previous attempts to improve QEC strategies have suffered from severe limitation regarding realistic setups, flexibility and scalability.
%
%In contrast to the approach of~\cite{FoeselTighineanuWeissMarquardt2018}, the method developed here can adapt and optimize QEC codes in terms of their resource efficiency, i.e., the number of physical qubits needed to achieve a desired maximal logical error rate, and operates without detailed information about or precise simulation of the underlying quantum states.
%%
%That is, our approach focuses on an adaptive code selection procedure that exploits available, sophisticated QEC techniques.
%In particular, we consider stabilizer codes~\cite{GottesmanPhD1997}, thereby significantly reducing the search space.
%
The proposed scheme can be employed both for off-line simulations with specified noise models and for on-line optimization for arbitrary, unknown noise, provided that potential hardware restrictions are taken into account. In particular, we demonstrate how the presented learning algorithm trained on a \emph{simulated} environment is able to \emph{transfer} its experience to different, physical setups. This \emph{transfer learning} skill is both remarkable and extremely useful: Maintaining the option to switch from one scenario to another, we can train a learning algorithm on fast simulations with modeled environments before optimizing the QEC code under real conditions. Transfer learning thus offers the ability to bootstrap the optimization of the actual quantum device via simulations. These simulations are comparably cheap since resources are much more limited under laboratory conditions. We show that our scheme is sufficiently efficient for %application to
setups expected to be available in the near-future~\cite{FriisMartyEtal2018, Zhang-Monroe2017, Bernien-Lukin2017}. In addition, these methods are parallelizable and hence expected to perform well also for larger system sizes. Our results thus suggest that RL is an effective tool for adaptively optimizing QEC codes.

\section{Framework \& Overview}

\begin{figure}[t!]
\begin{center}
	\label{fig:agent-environment}%(Color online)
	%%%trim={<left> <lower> <right> <upper>}
	\includegraphics[width=0.47\textwidth,trim={0mm 0mm 0mm 0mm}]{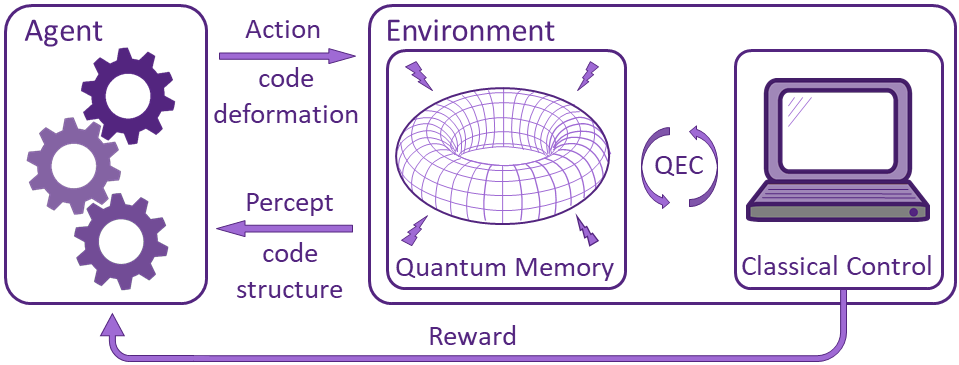}
	%\vspace*{-2mm}
	\caption{
		\textbf{Adapting quantum memories via reinforcement learning:} In the RL paradigm, an agent interacts with an environment through actions and percepts. The environment consists of a surface code quantum memory subject to noise, and a classical control system that guides the quantum error correction (QEC) on the memory and estimates logical error rates. The topological quantum memory embedded on a torus can be adapted through code deformations as instructed by the agent. In turn, the agent receives perceptual input in form of the current code structure and a reward which is issued by the classical control if the logical error rate is estimated to be below a desired threshold.}
\end{center}
\vspace*{-4mm}
\end{figure}

We consider a situation where a learning algorithm \textemdash referred to as an `agent' %in machine learning terminology
\textemdash interacts with an `environment' by modifying a given QEC code based on feedback from the environment. As illustrated in Fig.~\ref{fig:agent-environment}, this environment consists of a topological quantum memory~\cite{Terhal2015} (subject to noise) and its classical control system guiding the QEC. In each round of interaction, the agent receives perceptual input (\emph{`percepts'}) from the environment, that is, some information about the current code structure is provided.
The agent is tasked with modifying the code to achieve a logical error rate below a desired threshold. To this end, the agent is able to perform certain \emph{actions} in the form of fault-tolerant local deformations~\cite{BombinMartinDelgado2009, BravyiKitaev1998} of the code. The agent is \emph{rewarded} if the logical qubits have been successfully protected, i.e., if the logical error rate drops below the specified target. The problem of adapting QEC codes thus naturally fits within the structure of RL. Here it is important to note that the agent optimizing the quantum memory %in Fig.~\ref{fig:agent-environment}
is oblivious to the details of the environment. In particular, the agent cannot discern whether the environment is an actual experiment or a simulation.

The quantum memory we consider is based on the surface code~\cite{DennisKitaevLandahlPreskill2002, BravyiKitaev1998}, one of the most promising candidates for practical QEC~\cite{FowlerMariantoniMartinisCleland2012}. In particular, this versatile code is easily adaptable to compensate for a wide range of different types of errors~\cite{FujiiTokunaga2012}. Accordingly, we consider independent and identically distributed (i.i.d.) as well as non-i.i.d. errors on data qubits. We assume that a logical qubit is initially encoded in an 18-qubit surface code that can be extended by the agent by adding up to 50 additional data qubits via fault-tolerant local deformations~\cite{BravyiKitaev1998}.
Our simulations assume arbitrary qubit connectivity which allows the most flexibility to illustrate our approach, but the method can be adapted to different topologies that one may encounter in a specific hardware, as we discuss in Sec.~\ref{sec:hardware}.
At the heart of the classical control system is the SQUAB algorithm~\cite{DelfosseIyerPoulin2016, SQUABweb} simulating an optimal decoder in linear-time. SQUAB returns an estimation of the logical error rate in the (simulated) QEC procedure. As a specific learning model we employ %the physically motivated approach called
Projective Simulation (PS)~\cite{BriegelDeLasCuevas2012}, which has been shown to perform well in standard RL problems~\cite{MautnerMakmalManzanoTierschBriegel2015, MelnikovMakmalDunjkoBriegel2015, MelnikovMakmalBriegel2018}, in advanced robotics applications~\cite{HanglUgurSzedmakPiater2016}, and recently PS has been used to design new quantum experiments~\cite{MelnikovPoulsenNautrupKrennDunjkoTierschZeilingerBriegel2017}.

%%%%%%%%%%%%%%%%%%%%%%%%%%%%%%%%%%%%%%%%%%%%%%%%%%%%%%%%%%%%%%%%%%%%%%%%%%%%%%%%%%%%%%%%%%%%%%%%%%%%%%%
\begin{smallbox}{Summary of main results}{fidel}
%\label{fig:fidel}
%\begin{large}\begin{center} Summary of main results \end{center}\end{large}
\textbf{Reinforcement Learning framework}
\begin{itemize}[leftmargin=5mm]
    \item{for optimizing and adapting QEC codes, using}
    \item{arbitrary topological QEC codes,}
    \item{arbitrary decoders \& noise models,}
    \item{adaptable to any optimizer (RL paradigm)}
    \item{implementable on arbitrary platforms,}
    \item{applicable off-line (simulation) \& in-situ}
    %\item{dynamical online adaption.}
\end{itemize}
\textbf{Simulations using}
\begin{itemize}[leftmargin=5mm]
    \item{surface code quantum memory}
    \item{up to 68 fully connected data qubits,}
    \item{optimal$^{1}$ linear-time decoder (SQUAB)}
    \item{Projective Simulation model for RL}
\end{itemize}
\textbf{Simulations demonstrate agent's ability}
\begin{itemize}[leftmargin=5mm]
    \item{to determine optimal QEC codes}
    \item{for simple standard noise channels}
    \item{as well as non-isotropic noise, and for}
    \item{transfer learning in changing environments}
\end{itemize}
\end{smallbox}
%%%%%%%%%%%%%%%%%%%%%%%%%%%%%%%%%%%%%%%%%%%%%%%%%%%%%%%%%%%%%%%%%%%%%%%%%%%%%%%%%%%%%%%%%%%%%%%%%%%%%%%
\blfootnote{$^{1}$The decoder is optimal for the simplified approximate error model that we use for faster estimation of the logical error rate, see Sec.~\protect\ref{sec:res_efficiency}.}
\addtocounter{footnote}{1}

Within this framework, we demonstrate that agents can learn to adapt the QEC code based on limited resources to achieve logical error rates below a desired value for a variety of different error models. Moreover, we show that agents trained in such a way perform well also for modified requirements (e.g., lower thresholds) or changing noise models, and are thus able to transfer their experience to different circumstances~\cite{Thrun1996, WeissKhoshgoftaarWang2016}. In particular, we find that beneficial strategies learned by agents in a simulated environment with a simplified error model that allows for fast simulation of QEC~\cite{DelfosseZemor2017} are also successful for more realistic noise models. This showcases that agents trained on simulations off-line could be employed to `jump start' the optimization of physical quantum memories on-line in real-time. One reason for this adaptivity is that the agent obtains no information about the specifics of the noise, i.e., the latter appears to the agent as a black box environment. This further implies that our approach is hardware-agnostic, i.e., it can be applied to the same problem on different physical platforms, including trapped ions~\cite{Schindler-Blatt2013}, superconducting qubits~\cite{Barends-Martinis2014}, or topological qubits~\cite{Karzig-Freedman2017}. Depending on the particular platform, on-line optimization may be limited by specific hardware constraints (e.g., nearest-neighbour interactions), restricting, e.g., the possible code deformations (see Sec.~\ref{sec:hardware}).
%requires some flexibility of the underlying architecture, i.e. interactions which are not restricted to nearest neighbours
Nevertheless, off-line optimization still remains a viable option even for architectures with severe restrictions.
%Moreover, modifications that include non-nearest neighbour interactions may even be beneficial not only for on-line optimization but also from a functional point of view}~\cite{BravyiKoenig2013, PastawskiYoshida2015}.
Finally, note that the presented adaptive QEC framework in Fig.~\ref{fig:agent-environment} goes beyond the particular code, decoder, noise model, and RL method used here, and hence offers potential for extensions in all of these regards.

The remainder of this paper is structured as follows. In Sec.~\ref{sec:aqm}, we briefly review the surface code in the stabilizer formalism~\cite{GottesmanPhD1997} with emphasis on its inherent adaptability. Sec.~\ref{sec:rl_PS} introduces the concept of RL and the specific algorithm used. In Sec.~\ref{sec:results} various different scenarios within our framework are investigated. In Sec.~\ref{sec:on-off-line}, we discuss the efficiency of our simulations
%in more detail
and analyze the prospect of using results obtained in a simulated environment for more realistic, experimental settings. We conclude with a summary and outlook in Sec.~\ref{sec:discussion}.

\section{Adaptable Quantum Memory}\label{sec:aqm}

\subsection{Surface Code Quantum Memory}
%intro: QM
The purpose of a quantum memory is to encode logical information in such a way that small enough perturbations do not change the stored information.
%This is achieved through QEC which is the process of monitoring the memory and removing any errors that are detected.
This is achieved through QEC, which can, in parts, be understood within the stabilizer formalism~\cite{GottesmanPhD1997}.
Although a detailed background in QEC is not necessary to understand the basic principle of our approach, which can be viewed as the problem of restructuring a lattice (see Sec.~\ref{sec:aqm_adapt}),
let us briefly illustrate the stabilizer formalism for the example of a surface code quantum memory~\cite{BravyiKitaev1998}.

%In the following, we will briefly illustrate this formalism for the example of a surface code quantum memory~\cite{BravyiKitaev1998}.

%%
\begin{figure}[ht!]
\begin{center}
	\vspace{0.5cm}
	\includegraphics[width=0.4\textwidth]{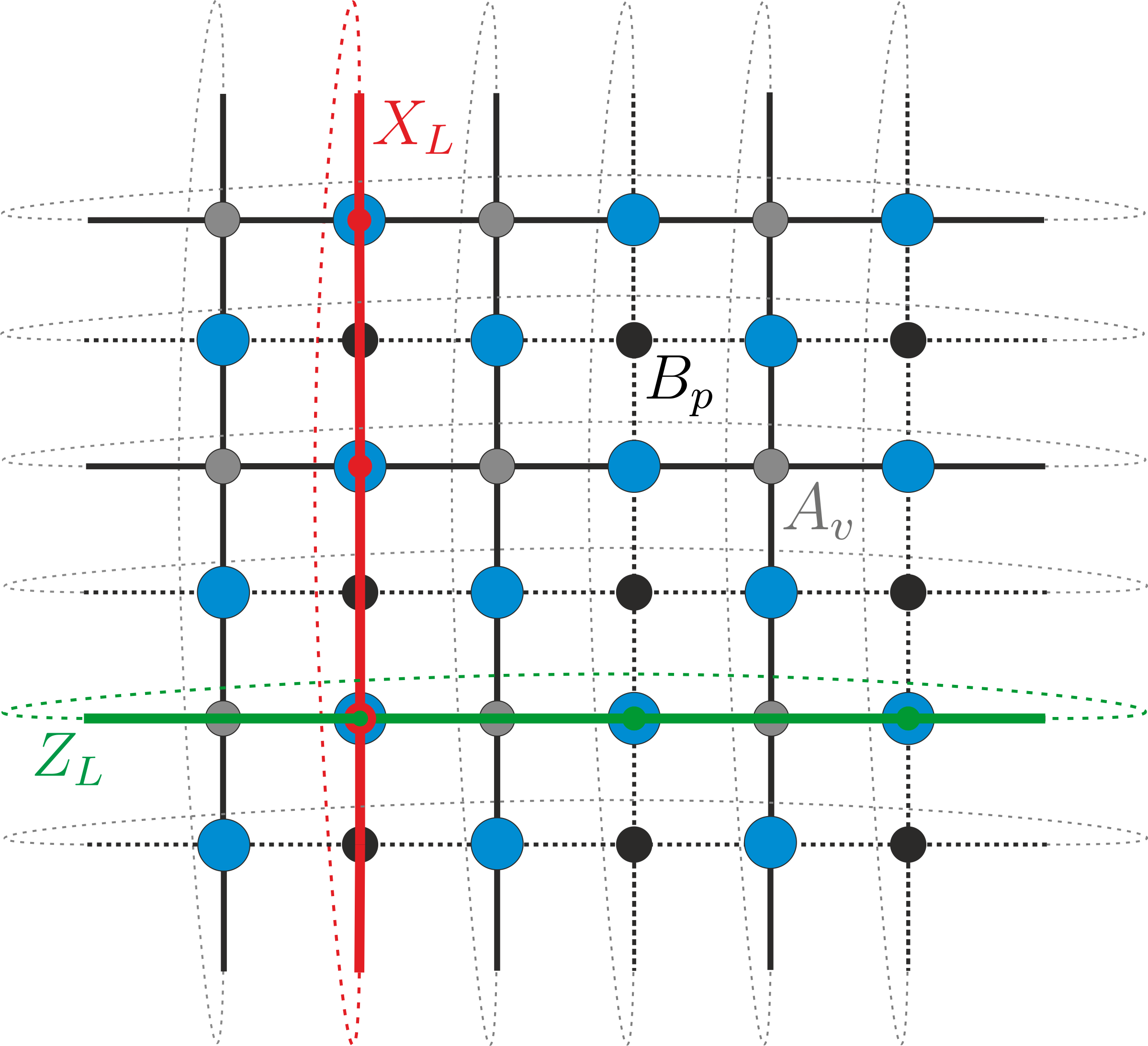}
	\caption{Standard surface code on a $3\times 3$ square lattice embedded on a torus. Larger, blue dots represent data qubits, smaller black and grey dots represent syndrome qubits. Lines connecting blue dots with vertices or plaquettes represent $Z$- and $X$-type stabilizers respectively. The black dotted line represents the dual lattice. The thick green line depicts a possible path for a $Z_\mathrm{L}$ string and the thick red line a path for a $X_\mathrm{L}$ string. This is also the initial code considered in Sec.~\ref{sec:results}.}\label{fig:sc}
\end{center}
\end{figure}
%%

%surface code + stabilizers
The surface code can be defined on any lattice embedded on a 2D manifold where data qubits are located on lattice edges (see Fig.~\ref{fig:sc}). These codes are called surface codes because each code in the class can be associated with a tessellation of a 2D manifold or surface~\cite{DennisKitaevLandahlPreskill2002}. For brevity, we consider lattices embedded on tori which define a subclass of surface codes, called toric codes~\cite{DennisKitaevLandahlPreskill2002}. However, the methods introduced in this paper can be easily extended to other surfaces or lattices with boundaries~\cite{FowlerMariantoniMartinisCleland2012}, and we hence just refer to surface codes in the following.
%This is also why we use both the term surface code, and toric code interchangeably.
Now, let us define products of Pauli operators for each vertex $v$ and plaquette $p$ as follows,
\begin{align}
A_v=\prod_{i\in \mathcal{N}(v)}X_i,\label{eq:scX}\\
B_p=\prod_{i\in \mathcal{N}(p)}Z_i.\label{eq:scZ}
\end{align}
Here, $\mathcal{N}(x)$ is the set of edges adjacent to a vertex $x=v$ or plaquette $x=p$ and $X_i$, $Z_i$ are Pauli-operators acting on the $i^{\mathrm{th}}$ qubit (indices enumerate data qubits located on the edges of the lattice).
We define the surface code's stabilizer group $\mathcal{S}$ as the group generated by all $A_v$ and $B_p$ under multiplication.
This group is Abelian, since all generating operators commute and have common eigenvectors. The $+1$ eigenspace defines the codespace $\mathcal{C}$ of the surface code. That is, a vector $\ket{\psi}\in\mathcal{C}$ iff $S\ket{\psi}=\ket{\psi}$ $\forall S\in \mathcal{S}$. Elements of $\mathcal{S}$ are called stabilizers.
Measuring the observables associated to these stabilizers allows checking whether the system state is still within the codespace.
%The weight of a stabilizer is the number of non-identity terms in the product.
%In order to encode $k$ logical qubits in $n$ physical qubits, $\mathcal{S}$ has to be generated by $s=n-k$ independent stabilizers. Hence, counting the number of faces and vertices versus the number of edges on a lattice embedded on a torus as in Fig.~\ref{fig:sc} gives the number of logical qubits in the surface code. That is, the surface code on a torus can encode two logical qubits.
For a surface code defined on a torus, the code space is isomorphic to $(\mathbb{C}^2)^{\otimes 2}$. It encodes $2$ logical qubits into $n$ data qubits where $n$ is the number of edges.

%logical operators
To each logical qubit one further associates a set of logical Pauli operators which define the centralizer of $\mathcal{S}$, i.e., operators that commute with $\mathcal{S}$. Trivial logical operators are stabilizers since they act as the identity on the logical subspace.
In the surface code, nontrivial logical operators are strings of tensor products of Pauli operators along topologically nontrivial cycles, i.e., noncontractible paths, of the torus. Each logical qubit is defined by a logical $Z$-operator along a noncontractible path of the lattice and a logical $X$ running along a topologically distinct noncontractible path of the dual lattice (see Fig.~\ref{fig:sc}). Since such paths necessarily cross an odd number of times, they anticommute.

%error correction + distance
Logical operators change the encoded information.
Thus, it is desirable to detect and correct errors before they accumulate to realize a logical operation.
Errors which are neither stabilizers nor logical operators can be detected by stabilizer measurements since they anticommute with some stabilizers. In order to perform stabilizer measurements, one first associates so-called syndrome qubits with vertices and faces. Then, the syndromes, i.e., the eigenvalues of stabilizers, are obtained by entangling syndrome qubits with adjacent data qubits and subsequent measurements~\cite{FowlerMariantoniMartinisCleland2012}. In a fixed architecture the required entangling gates between syndrome and data qubits may be represented by the edges of the lattice (for $X$-stabilizers) and its dual (for $Z$-stabilizers).

The length of the  shortest path of any noncontractible loop on the torus is the distance $d$ of the surface code. That is, the distance is the minimal number of physical errors that need to occur in order to realize a logical operator. Any number of errors below $d$ can be detected but not necessarily corrected and less than $d/2$ errors can always be corrected in the stabilizer formalism~\cite{GottesmanPhD1997}.
% keep remove?
%To see this, note that stabilizer measurements can detect endpoints of strings of errors. The aim of QEC is to close open ended strings such that the resulting operator acts trivially on the encoded subspace. That is, strings have to be corrected to a stabilizer element or identity. In the surface code, stabilizers can be identified with trivial cycles on the lattice. A string of errors that closes a nontrivial cycle halfway (acting on at least $d/2$ qubits) will likely be closed incorrectly and cause a logical errors.

\subsection{Adaptable Surface Code}\label{sec:aqm_adapt}
%adaptiveness
So far, we have assumed the standard notion of a surface code defined on a square lattice on a torus (see Fig.~\ref{fig:sc}). However, surface codes can be defined on arbitrary lattices. Since the surface code stabilizers in Eqs.~(\ref{eq:scX}) and~(\ref{eq:scZ}) are identified with vertices and plaquettes of the lattice, adapting the lattice can change the underlying code and its performance dramatically~\cite{FujiiTokunaga2012}. This feature makes surface codes particularly useful for biased noise models, e.g., noise models where $X$- and $Z$-errors occur with different probability. Even the loss tolerance is affected by the underlying lattice.
%This bias in error correction comes from error decoding where it is less likely to correctly reconstruct error paths on a lattice with high connectivity. For instance, a dual lattice with high connectivity indicates that the average number of multipliers making up $Z$-type stabilizers (see Eq.~(\ref{eq:scZ})) is high.
%Besides the connectivity, the number of vertices as compared to the number of plaquettes may change depending on the lattice. That is, the number of $X$- and $Z$-stabilizer can differ from surface to surface code.

%changing lattice
In our framework, we will be exploring the space of lattices to make use of the adaptability of the surface code. To search this space in a structured fashion, we introduce a set of elementary lattice modifications.

\begin{figure}[t!]
\begin{center}
	\vspace{0.5cm}
	\subfigure[]{\includegraphics[width=0.4\textwidth]{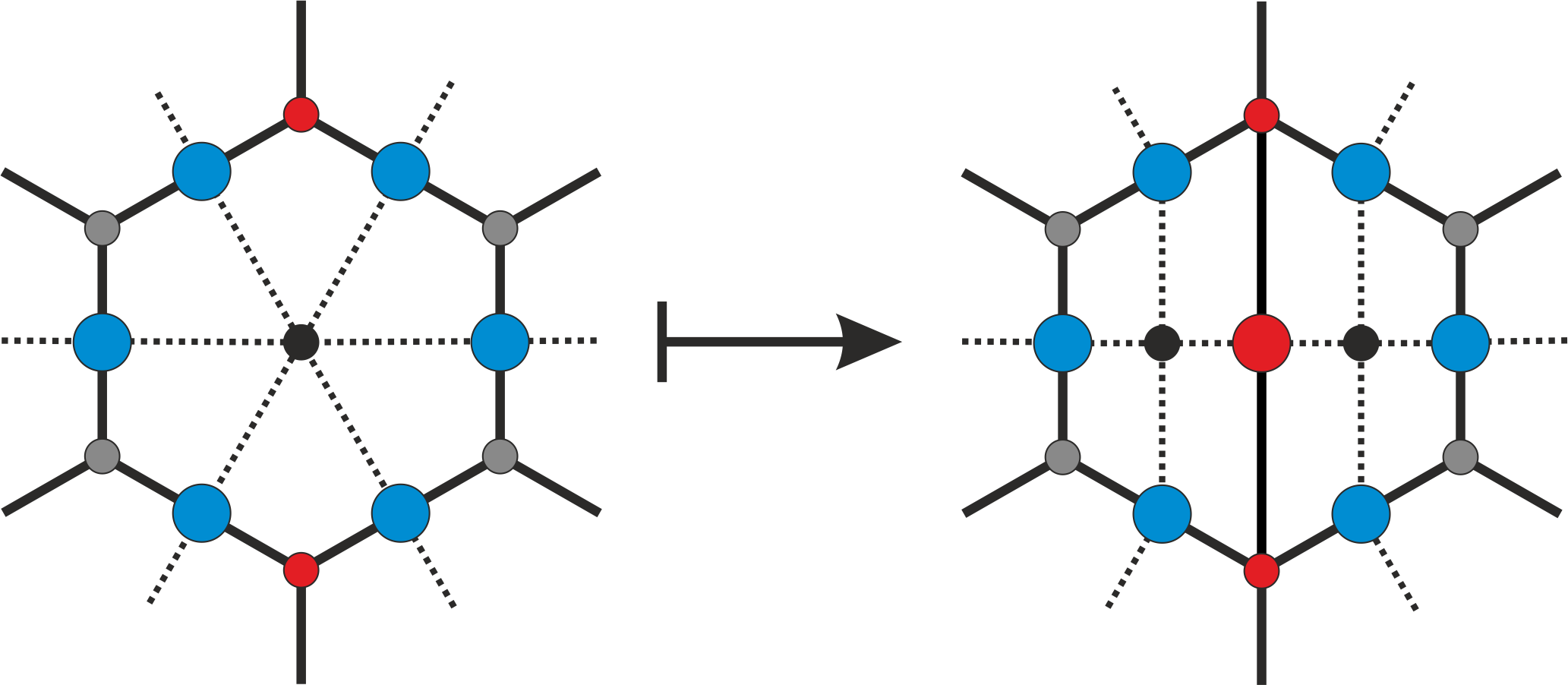}
	\label{fig:mva}}\hspace{0.30cm}
	\subfigure[]{\includegraphics[width=0.4\textwidth]{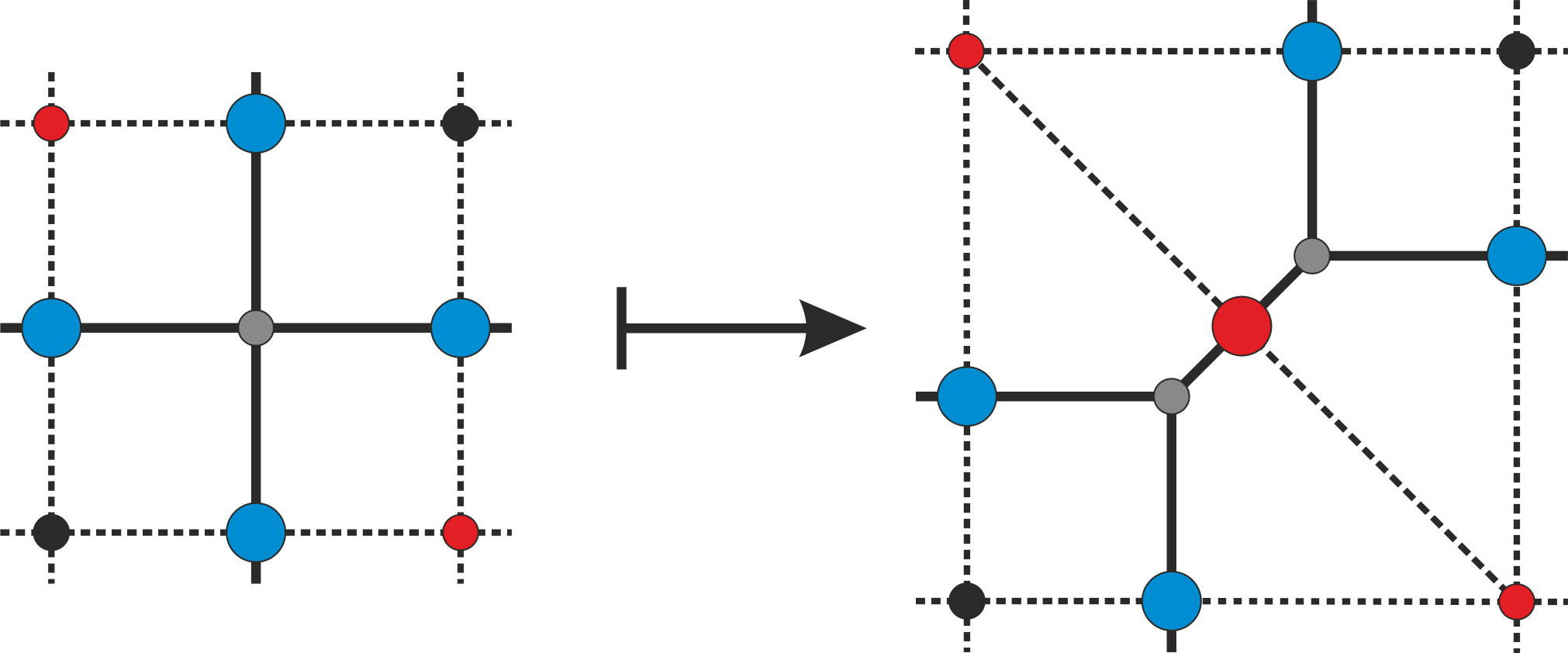}
	\label{fig:mvb}}
	\vspace*{-3mm}
	\caption{An illustration of the basic moves that fault-tolerantly map surface codes to surface codes while changing the underlying lattice and hence, the stabilizer group. (a) Two vertices (here chosen at the top and bottom in red) are connected across a plaquette (central black dot) which is effectively split into two by a new edge. (b) Two plaquettes (red dots on top left and bottom right) are connected across a vertex (central grey dot) which is effectively split into two by a new edge. In both cases the number of data qubits (blue dots) is increased by one (central red) qubit. Note that move (b) can be understood as a move of type (a) on the dual lattice and vice versa.}\label{fig:mv}
\end{center}
\end{figure}

%In order to adapt the surface code, the underlying lattice has to be changed fault-tolerantly. Here, fault-tolerance requires changes to be local w.r.t. the lattice topology. More specifically, the circuit that implements the changes has to act on a small number of physical qubits. In this way, errors can spread only to a limited number of qubits even if each gate in the circuit fails.
Specifically, we consider modifications that can be performed with a set of basic moves which were first illustrated in Ref.~\cite{BravyiKitaev1998}. The two basic moves are exemplified in Fig.~\ref{fig:mv}: (i) As illustrated in Fig.~\ref{fig:mv}(a), addressing the primal lattice, two non-neighbouring vertices can be connected across a plaquette which is effectively split by the additional edge. That is, the number of $Z$-stabilizers is increased while an additional data qubit is added. (ii) Conversely, the same operation can be performed on the dual lattice producing an additional vertex (i.e., $X$-stabilizer) as illustrated in Fig.~\ref{fig:mv}(b). Note that the reverse operation, i.e., deleting an edge by merging two vertices or two faces, respectively, is also possible. While the inclusion of qubit deletion greatly increases the number of possible moves (actions) in each step, the set of reachable codes is the same as if one had started from a single data qubit using only the operations in Fig.~\ref{fig:mv}. Here, we have chosen not to explicitly include such deletions to simplify the simulations. All of these moves can be implemented fault-tolerantly \footnote{Here, fault-tolerance requires changes to be local w.r.t. the lattice topology. More specifically, the circuit that implements the changes has to act on a small number of data qubits. In this way, errors can spread only to a limited number of qubits even if each gate in the circuit fails.} and map a surface code to a surface code while changing the underlying stabilizer group~\cite{BravyiKitaev1998}. This can be considered as simple version of code deformation~\cite{BombinMartinDelgado2009} which is employed here to explore the space of surface codes.
Indeed, code deformation can in principle be employed in a much more general way to map topological stabilizer codes to other topological stabilizer codes.
However, for our needs the two basic moves explained above already allow for sufficiently diverse modifications while limiting the number of available actions. Nevertheless, note that the number of available moves increases with the lattice size.

\subsection{Hardware constraints}\label{sec:hardware}

\begin{figure*}[ht!]
\begin{center}
	\subfigure[]{\includegraphics[width=0.22\textwidth]{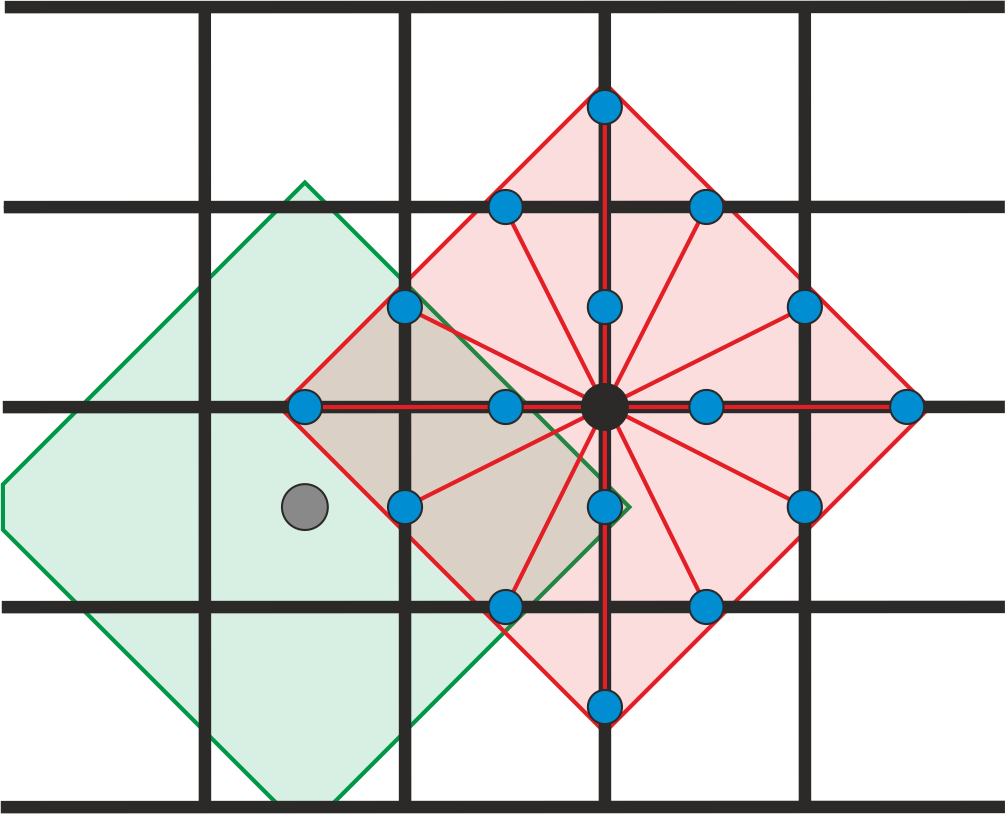}
		\label{fig:2NNConnections}}\hspace{0.30cm}
	\subfigure[]{\includegraphics[width=0.22\textwidth]{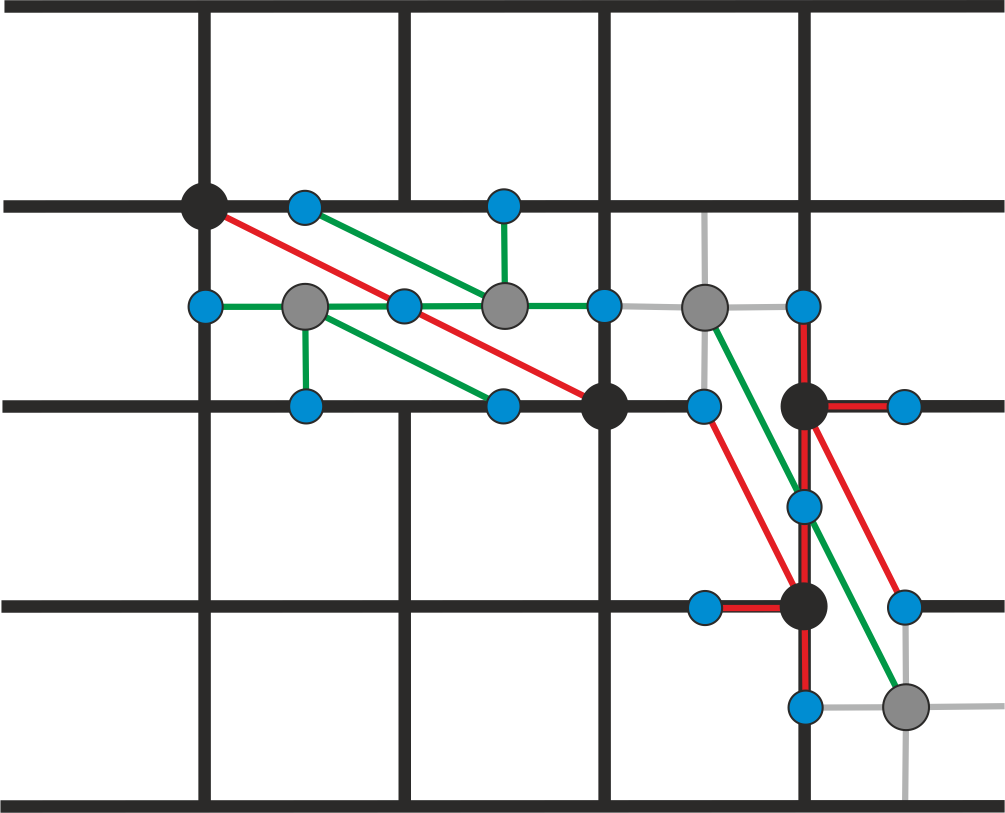}
		\label{fig:2NNConnectivity}}\hspace{0.30cm}
	\subfigure[]{\includegraphics[width=0.22\textwidth]{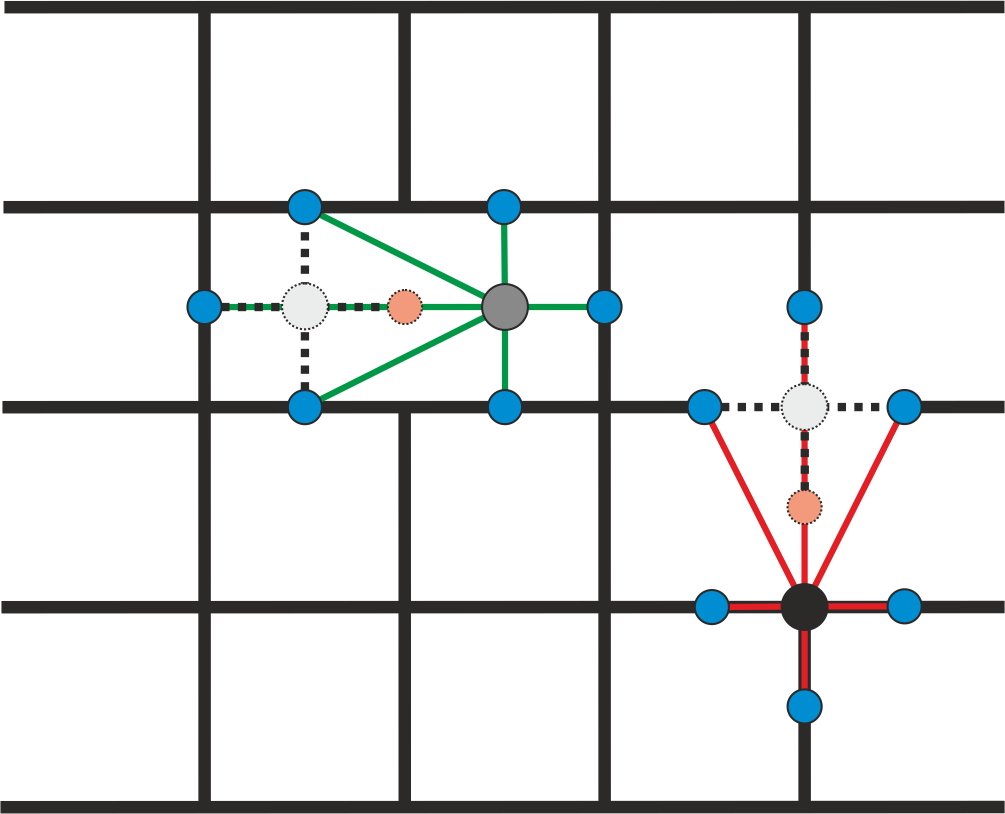}
		\label{fig:2NNDataRM}}\hspace{0.30cm}
	\subfigure[]{\includegraphics[width=0.22\textwidth]{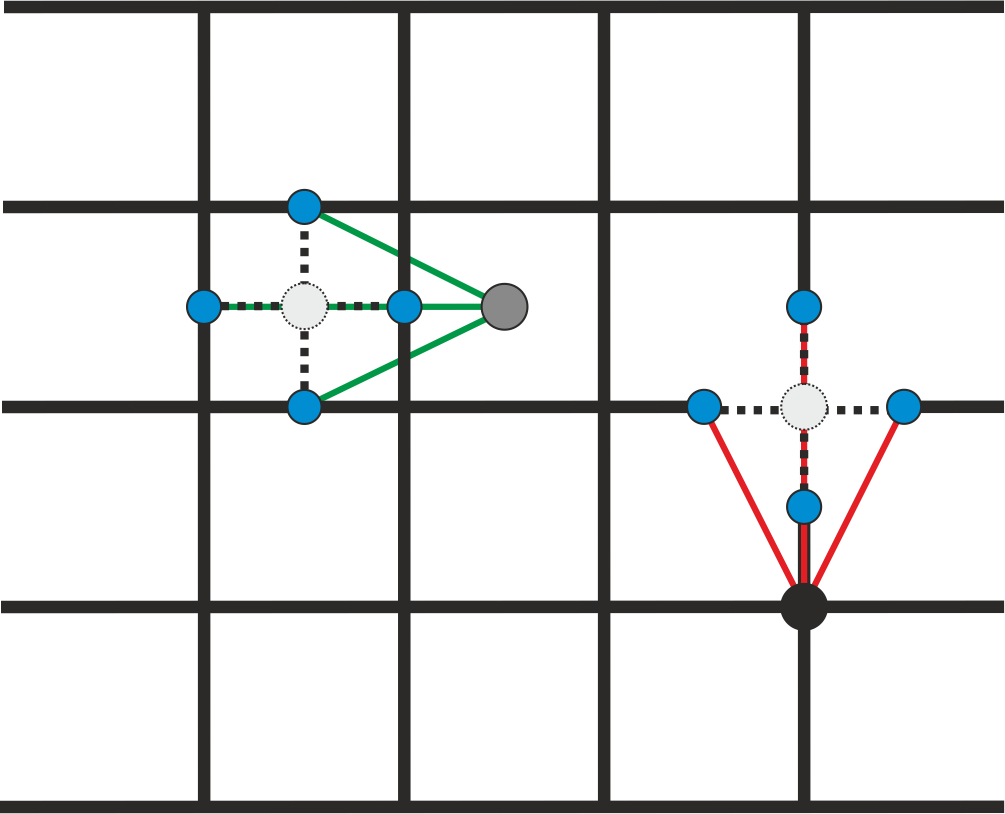}
		\label{fig:2NNSyndromeRM}}
	\vspace*{-3mm}
	\caption{%(Color online)
	A selection of possible lattice configurations for a 2-nearest neighbour surface code architecture with boundaries illustrating the available flexibility. Blue dots represent data qubits, black and grey dots represent syndrome qubits.  Connecting lines visualize potential vertex and plaquette
    stabilizers. (a) A 2-nearest neighbour surface code on a square lattice with additional connections shown. For each vertex and plaquette the possible connections (for syndrome readouts) extend within a square of size two as displayed for one vertex and one plaquette. (b) Exemplary lattice with changed connectivity. (c)  Lattice configuration where two data qubits (red dots) and two syndrome qubits (white dots) have been removed. (d) In a 2-nearest neighbour setup a syndrome qubit and its connections can be replaced by a neighbouring syndrome qubit. Dotted lines represent previous connections to removed qubits that are no longer used.
	}
	\label{fig:2NNHardware}
\end{center}
\end{figure*}

Before we further go into detail on our numerical study focussing on setups with the full flexibility required to implement the moves in Fig.~\ref{fig:mv}, let us briefly consider the implications of the variations in hardware of currently available setups. For instance, limitations on how freely qubits that are physically present but not yet part of a specific code structure can be connected to the latter vary between different platforms such as ion trap architectures~\cite{Schindler-Blatt2013, Amini-Wineland2010, Bowler-Wineland2012} or superconducting qubit devices~\cite{Barends-Martinis2014}. While our method can be used directly when all-to-all connectivity between qubits is given, e.g., in many current ion traps~\cite{Schindler-Blatt2013, Amini-Wineland2010, Bowler-Wineland2012}, modifications can be made to the set of allowed code deformations when this is not the case, provided that some minimal requirements are met. That is, for any fixed architecture to be able to compensate for biased and potentially correlated noise it is required that (i) the connectivity of the lattice can be changed, (ii) data qubits can be removed and (iii) syndrome qubits can be removed. While changing the connectivity of the lattice enables one to adapt to biased error channels in accordance with Ref.~\cite{FujiiTokunaga2012}, removing qubits allows the adaptability to correlated noise sources arising, e.g., from manufacturing defects. In very restricted scenarios, for example, when only nearest neighbour interactions are possible, these requirements are not met. In such cases, one may still perform code optimization off-line using an estimated noise model in order to produce a blueprint for manufacturing or initialization. However, already a 2-nearest neighbour architecture (featuring connections between syndrome qubits and next nearest data qubits) provides the required features, as can be seen from Fig.~\ref{fig:2NNHardware}.

At this point, it should also be mentioned that the flexibility provided by higher connectivity in terms of code adaptability is not the only relevant factor: Higher connectivity might be subject to a trade-off with lower fidelities due to higher technical demands. At the same time, allowing for flexible code deformations and adaptive on-line error correction is certainly not the only motivation for non-nearest neighbour connections. In particular, nearest neighbour architectures may suffer from limitations on fault-tolerant implementations of logical gates~\cite{BravyiKoenig2013,PastawskiYoshida2015}. Indeed, the first successful experiments with a non-nearest neighbour architecture of superconducting qubits have been successfully performed~\cite{Rosenberg-Oliver2017}, suggesting that (variations of) our methods are applicable across a range of platforms currently in use or under development. Despite the hardware constraints, we still restrict our actions to fault-tolerant, local deformations since this provides a meaningful way of exploring the family of surface codes in the context of RL for in-situ optimization.

\section{Reinforcement Learning and Projective Simulation}\label{sec:rl_PS}
%RL + transfer learning
In generic RL settings~\cite{SuttonBarto1998}, one considers an agent that is interacting with an environment (see Fig.~\ref{fig:agent-environment}). At each time step the agent obtains information from the environment in terms of perceptual input -- percepts -- and responds through certain actions. Specific behavior, i.e., sequences and combinations of percepts and actions, trigger reinforcement signals -- rewards. Typically, the agent adapts its behavior to maximize the received reward per time step over the course of many rounds of interactions.
%In a generic RL setting, one considers an agent that is situated in an environment (see Fig.~\ref{fig:agent-environment}). The agent interacts with the environment through \textit{actions} while the environment responses with reinforcement signals -- rewards -- which reinforce certain beneficial behaviors. The current state of the environment is perceived by the agent through \textit{percepts}. The agent's task is to maximize the received reward through many interactions with the environment.
%Stated this way, RL is a very generic approach to problem solving.
The RL method as we use it makes no assumptions about the environment and applies even if the environment is a black box. In our case, the environment is a surface code memory subject to noise and its classical control system, including the means to estimate logical error rates. Here, knowledge of the environment is also restricted: The agent is never directly provided with information about the noise model. This is because, in practice, acquiring knowledge about the noise requires involved process tomography.
Since the agent does not aim to characterize the noise process, but rather to alleviate its effects, a complicated noise estimation procedure is not necessary.
%At the same time, exploring the environment, an RL algorithm slowly stores knowledge about the underlying search space itself~\cite{MelnikovPoulsenNautrupKrennDunjkoTierschZeilingerBriegel2017}. This knowledge can -- in principle -- be transferred to similar tasks. When an RL algorithm successfully employs knowledge between different tasks or environments, it is called transfer learning~\cite{Thrun1996}. This can yield a significant reduction in learning time since the agent is not required to start from scratch every time the task or environment changes slightly~\cite{TommasiCaputo2009,TommasiOrabonaCaputo2010,AytarZisserman2011}. Transfer learning is an active area of research in general AI~\cite{WeissKhoshgoftaarWang2016}.

\begin{figure}[t!]
\begin{center}
	\vspace{0.5cm}
	\includegraphics[width=0.46\textwidth]{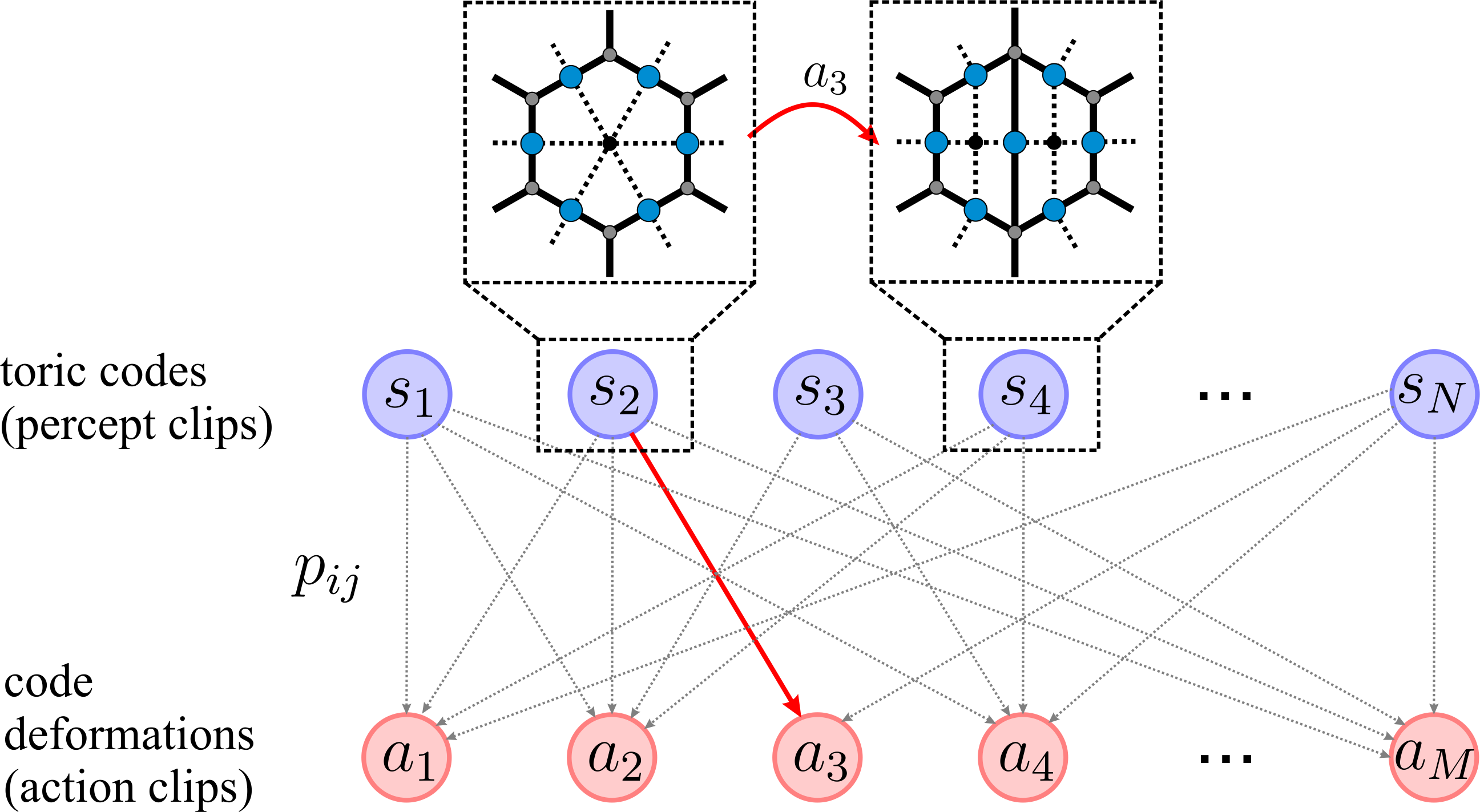}
	\caption{Two-layered clip network of a PS agent. Percept clips in the first (upper) layer are colored blue and represent surface codes. Action clips are colored red and represent code deformations. Edges $(i,j)$ carry weights that can be associated with transition probabilities $p_{ij}$ in accordance with Eq.~(\ref{eq:transitionprob}). If an action, say $a_3$, is performed upon seeing a percept, say $s_2$, the code associated with $s_2$ is adapted according to $a_3$. This results in a new code, say $s_4$, that is perceived by the agent. Note that not every percept is necessarily connected to every action since the set of available moves in Fig.~\ref{fig:mv} depends on the lattice, i.e. the percept.}\label{fig:network}
\end{center}
\vspace*{-2mm}
\end{figure}

%PS
The specific learning agent that is considered in this paper is based on the Projective Simulation (PS) model for RL. PS is a physics-motivated framework for artificial intelligence developed in Ref.~\cite{BriegelDeLasCuevas2012}.
%It has been shown to perform well in standard RL problems~\cite{MautnerMakmalManzanoTierschBriegel2015, MelnikovMakmalBriegel2018, MelnikovMakmalDunjkoBriegel2015},  in advanced robotics applications~\cite{HanglUgurSzedmakPiater2016} and recently PS has been used to design new quantum experiments~\cite{MelnikovPoulsenNautrupKrennDunjkoTierschZeilingerBriegel2017}.
%In the following, we will describe the basic mechanism for PS in terms of the generic RL setting developed above.
The core component of a PS agent is its clip network which is comprised of units of episodic memory called \textit{clips}~(see Fig.~\ref{fig:network}). There are two sets of clips constituting the basic network, \textit{percept} and \textit{action} clips. In an interactive RL scenario, an agent \textit{perceives} the environment through the activation of a percept clip $s_i\in P$ and responds with an \textit{action}. The latter, in turn, is triggered by an action clip $a_j\in A$.
%If a percept issued by the environment has not been seen before, a new clip is automatically created in the clip network.
Percept clips can be regarded as representations of the possible states of the environment, as perceived by the agent.
Similarly, action clips can be seen as internal representations of operations an agent can perform on the environment.
A two-layered clip network as in Fig.~\ref{fig:network} can be represented by a directed, bipartite graph where the two disjoint sets comprise the percepts $P$ and actions $A$, respectively.
In this network each percept (clip) $s_i\in P$, $i \in[1,N^{(t)}]$ (where $N^{(t)}$ is the number of percepts at time step $t$) is connected to an action (clip) $a_j\in A$, $j\in[1,M_i]$ (with $M_i$ being the number of actions available for a percept $s_i$) via a directed edge $(i,j)$ which represents the possibility of taking action $a_j$ given the situation $s_i$ with probability $p_{ij}:=p(a_j|s_i)$.
The agent's policy governing its behavior in this RL setting is defined by the transition probabilities in the episodic memory.
Learning is manifested through the adaption of the clip network via the creation of new clips and the update of transition probabilities. Each time a new percept is encountered, it is included into the set $P$. A time-dependent weight, called $h$-value, $h_{ij}^{(t)}$ is associated with each edge $(i,j)$. The transition probability from percept $s_i$ to action $a_j$ is given by the so-called softmax function~\cite{SuttonBarto1998} of the weight $\sigma_\beta(h_{ij}^{(t)})$,
\begin{align}
p^{(t)}_{ij}=
\frac{e^{\beta\hspace{0.05em} h_{ij}^{(t)}}}{\sum_{k=1}^{M_i}e^{\beta\hspace{0.05em} h_{ik}^{(t)}}},\label{eq:transitionprob}
\end{align}
where $\beta>0$ is the softmax parameter.

When a percept is triggered for the first time all $h$-values are set to~$1$ such that the transition probabilities are initially uniform. That is, the agent's behavior is fully random in the beginning. Since random behavior is rarely optimal, changes in the transition probabilities are required. The environment reinforces such changes by issuing nonnegative rewards $\lambda^{(t)}$ in response to an action of the agent. Then, the agent must ask the question, given a percept $s_i$ which is the action $a_j$ that will maximize the received reward. Therefore, the transition probabilities are updated in accordance to the environment's feedback such that the chance of the agent to receive a reward in the future is increased. In other words, the environment's feedback $\lambda^{(t)}$ controls the update of the $h$-matrix with entries $h_{ij}$. However, there are many other contributions to the update independent of the environment's immediate feedback. Particularly noteworthy are contributions that reinforce exploratory over exploitative behavior. A detailed description of how the feedback is processed in the agent's memory is given in Appendix~\ref{app:PS}.

Once an agent has learned to optimize the reward per interaction step, the clip network in Fig.~\ref{fig:network} is a
representation of the agent's policy: Nonuniform transition probabilities $p_{ij}$ mark a systematic decision-making aimed at maximizing the reward per time step while accounting for exploratory behavior. Hence, the network can also be interpreted as an agent's memory encoding knowledge of the reward landscape of the environment.

\section{Optimizing Quantum Memories - a reinforcement learning problem}\label{sec:results}
%%
%RL and QEC
In this section, we show that RL can be employed within the framework of Fig.~\ref{fig:agent-environment} to successfully optimize QEC codes w.r.t. their resource efficiency. In our framework for adaptive quantum error correction, we consider a PS agent that interacts with a specific, simulated environment -- the surface code quantum memory subject to noise and its classical control managing the QEC procedure. We use the term environment in the sense it is used in RL rather than referring to the source of noise as is common in, say, open-system quantum physics.
%The agent acts on the environment through actions which constitute changes in the code structure while the environment responses with reinforcement signals. Reinforcement signals are rewards issued by the environment to reinforce the agent's behavior towards a low logical error rate. The current state of the environment as seen by the agent through percepts is represented by the current code structure. The details of the algorithm, specifically the environment, are presented in Appendix~\ref{app:env} and the clip network is visualized in Fig.~\ref{fig:network}.

%Specific scenario under consideration
To be more precise, we start each trial by initializing a distance-three surface code quantum memory with 18 qubits (edges) defined on a square lattice $\Lambda$ embedded on a torus (see Fig.~\ref{fig:sc}).
Note that the choice of initial code is ad hoc and could be replaced by any other small-scale surface code designed to suit a given architecture.
The code is subject to an unknown Pauli noise channel $\mathcal{E}$ which may change in time and may differ for different data qubits. The classical control simulates the QEC procedure under this noise model and estimates the logical error rate $P_\mathrm{L}$. Having a basic set of moves at its disposal (see Fig.~\ref{fig:mv}), the agent is tasked with growing the lattice until the logical error rate is below a target $P_\mathrm{L}^{\mathrm{rew}}$ or at most $50$ additional qubits have been added. This choice of an upper limit should be viewed in light of recent progress in the development of quantum devices with more than $50$ qubits~\cite{Zhang-Monroe2017, Bernien-Lukin2017}. Note the difficulty of the simulation task. A single trial requires to simulate the perfomance of up to $50$ QEC codes.
%while never exceeding an intermediate target $P_\mathrm{L}^{\mathrm{int}}$.
The basic set of moves is rather generic and could be restricted further to suite the requirements of a given hardware. For instance, actions could be restricted to the boundary of a system.
Once the desired logical error rate is reached, the agent receives a reward $\lambda=1$, the trial ends and the quantum memory is reset to $\Lambda$. If the desired logical error rate is not reached before $50$ qubits have been introduced, the code is reset without reward.
%If the logical error rate at any timestep $t$ exceeds $P_\mathrm{L}^{\mathrm{int}}$, the code is reset without reward.
The details of the algorithm, specifically the environment, are presented in Appendix~\ref{app:env} and the agent's clip network, i.e., episodic memory, is visualized in Fig.~\ref{fig:network}.
Note that this scenario -- although restricted to surface codes -- is already extremely complex and versatile. In Appendix~\ref{app:search},
we analyze the difficulty of this problem and show that there is no hope of solving it by random search.

\begin{figure*}[ht!]
\begin{center}
	\subfigure[]{\includegraphics[width=0.48\textwidth]{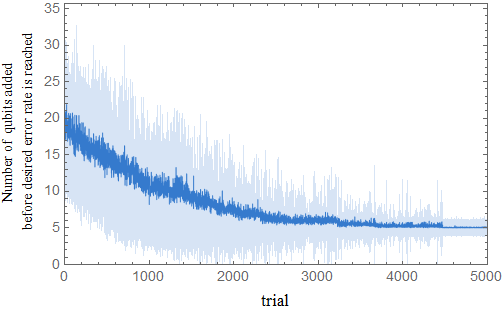}
		\label{fig:resultiidZ}}\hspace{0.30cm}
	\subfigure[]{\includegraphics[width=0.44\textwidth]{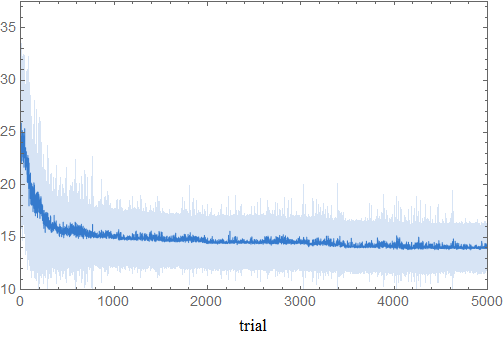}
		\label{fig:resultiidXZ}}
	\vspace*{-3mm}
	\caption{%(Color online)
		The figure shows the results for two simple tasks as considered in Sec.~\ref{sec:res_iid} which confirm that RL learning can be applied to the problem of optimizing quantum memories. The area shaded in light blue is the standard deviation over 60 agents. Given the available set of actions, shown in Fig.~\ref{fig:mv}, the PS agent can add, one-by-one, qubits to the initial code (see Fig.~\ref{fig:sc}) until the logical error rate is below $P_\mathrm{L}^{\mathrm{rew}}=0.001$. (a) The memory is subjected to the noise channel in Eq.~(\ref{eq:errorZ}) with $p=0.1$. The initial logical error rate is $P_L^{\textrm{init}}\approx0.006$. The number of qubits to reach the desired error rate $P_\mathrm{L}^{\mathrm{rew}}$ per trial is depicted. (b) The same task is shown with the error channel of Eq.~(\ref{eq:errorXZ}) such that $P_L^{\textrm{init}}\approx0.005$.
		Different learning parameters (see Appendix~\ref{app:PS}) are reflected in different learning times, standard deviations and convergence behavior. Generally, there is a trade off between learning time and standard deviation as can be seen from the comparison between (a) and (b). Shaded areas represent the standard deviation. Details about parameters are given in Appendix~\ref{app:param}.
	}
	\label{fig:resultiid}
\end{center}
\end{figure*}
%%

%outlook
In the following, we verify that the task outlined above is indeed an RL problem where learning is beneficial. To this end, we start by considering some instructive cases where the solutions are known before moving on to a more complicated scenario with unknown optimal strategy. The variables that are being changed in between scenarios are the error channel $\mathcal{E}$ or the rewarded logical error rate $P_\mathrm{L}^{\mathrm{rew}}$.
However, if not stated otherwise, we set $P_\mathrm{L}^{\mathrm{rew}}=0.001$.
For illustrative reasons, the error models $\mathcal{E}$ that appear in the following are represented by single-qubit Pauli error channels. In fact,
	%this channel will just serve as a motivation to construct a
the simulated quantum memory will be subject to a so-called erasure channel~\cite{BennettDiVincenzoSmolin1997, GrasslBethPellizzari1997} which models the behavior of the actual Pauli channel. This particular choice of error model is motivated by the availability of an optimal decoder for erasure noise on a surface code~\cite{DelfosseZemor2017}. In Sec.~\ref{sec:res_efficiency}, this simplified channel is analyzed in more detail and we verify that it is suitable to model actual error channels.
\subsection{QEC codes for i.i.d. error channels}\label{sec:res_iid}
\begin{figure}[t!]
\begin{center}
	\vspace{0.5cm}
	\includegraphics[width=0.4\textwidth]{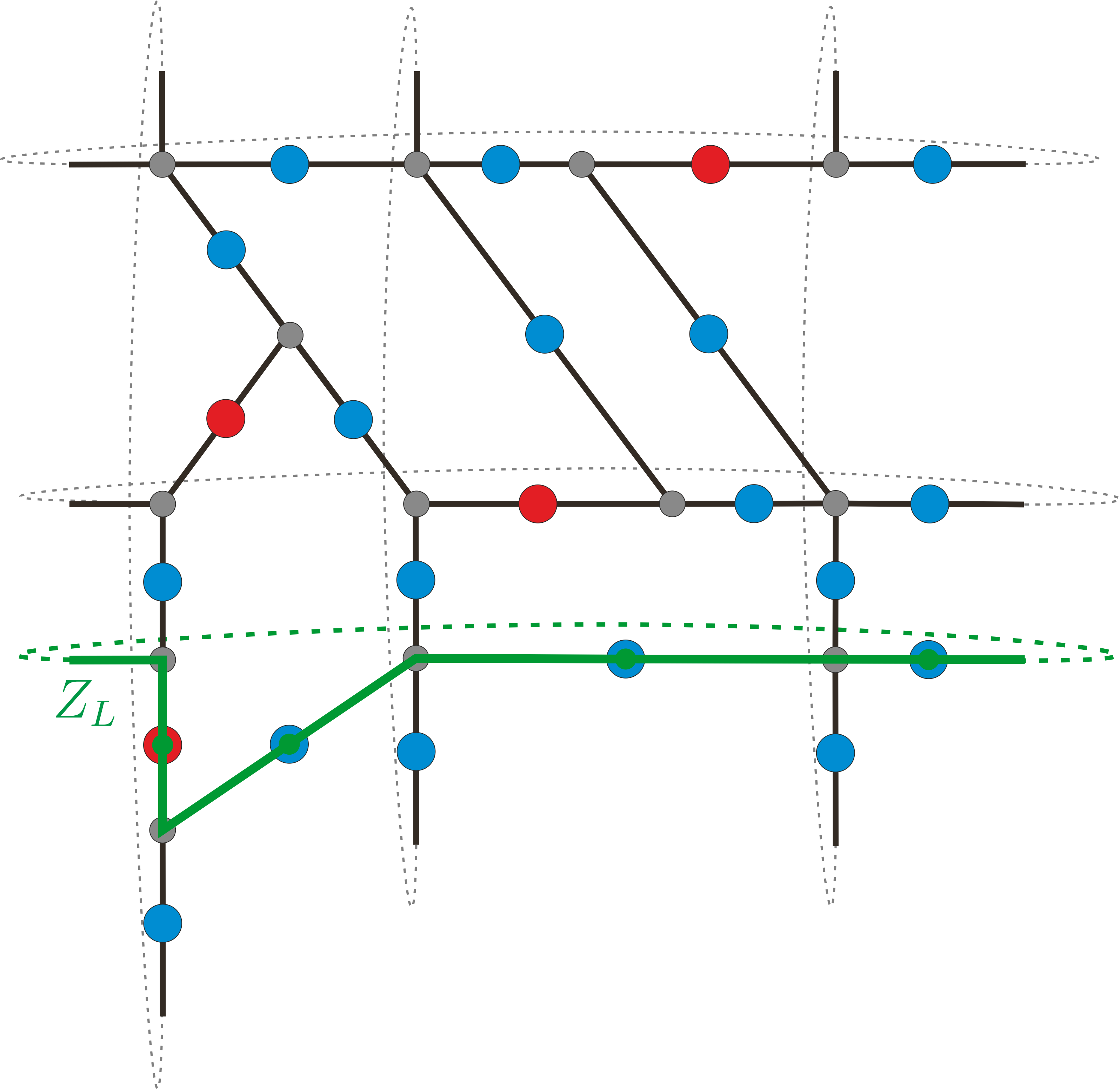}
	\caption{
		The best agents only require $4$ additional qubits to solve the task described in Sec.~\ref{sec:res_iid} with the error channel in Eq.~(\ref{eq:errorZ}). Here, an exemplary solution is shown, i.e., a surface code with a logical error rate below $P_\mathrm{L}^{\mathrm{rew}}=0.001$. Red circles represent qubits added to the code. Removing all red qubits, we recover the initial code (see Fig.~\ref{fig:sc}). As expected, all actions increase the number of $X$-stabilizers in order to protect against $Z$-errors. The thick green line depicts a possible path for a $Z_\mathrm{L}$ string crossing~$4$ qubits.
	}\label{fig:graphiidZ}
\end{center}
\vspace*{-3mm}
\end{figure}

As a first simple task, we consider two straightforward scenarios with known good solutions as they will nonetheless be crucial in the evaluation of the behavior of the RL agent.
First, let us consider an error channel that can only cause $Z$-errors with probability $p=0.1$ on each individual, data qubit independently. Formally, we can write this as a quantum channel acting on each single-qubit state $\rho$ as,
\begin{align}
\mathcal{E}(\rho)=p\: Z\rho Z+(1-p)\rho.\label{eq:errorZ}
\end{align}
Assuming the initial logical error rate is above the target value $P_\mathrm{L}^{\mathrm{rew}}$, the RL agent is then tasked with modifying the 18-qubit surface code by increasing the number of data qubits according to the rules described in Fig.~\ref{fig:mv}. At the beginning of each trial, an agent has a reservoir of 50 additional qubits at its disposal to reduce the logical error rate below $P_\mathrm{L}^{\mathrm{rew}}=0.001$.
Fig.~\ref{fig:resultiidZ} shows that the agent indeed learns a strategy, i.e., a policy, which adds -- on average -- $5$ data qubits to the initial code in order to reduce the logical error rate as desired.
Fig.~\ref{fig:resultiidZ} can be understood as follows. In the very beginning, the agent has no knowledge about the environment and performs random modifications of the surface code. For the specific error channel in Eq.~(\ref{eq:errorZ}), a random strategy requires on average 20 additional data qubits to reach the desired logical error rate $P_\mathrm{L}^{\mathrm{rew}}$.
What follows are many trials of exploring the space of surface codes. During each trial, the agent's episodic memory is reshaped and modified as described in Sec.~\ref{sec:rl_PS} and Appendix~\ref{app:PS}. This learning process effectively increases the probability for the agent to select sequences of actions which lead to a reward quickly. As can be seen from Fig.~\ref{fig:resultiid} the length of a rewarded sequence of actions is gradually reduced with each trial. Ideally, the agent's behavior converges to a policy that requires a minimum number of additional qubits.
In Fig.~\ref{fig:resultiidZ} we observe that agents converge towards a policy which requires, on average, $5$ additional qubits.
Let us now confirm that the strategy found by the RL agents agrees with the best known strategies for this problem.
The error channel in Eq.~(\ref{eq:errorZ}) can only cause logical $Z$-errors. That is, we are looking for the surface code with the maximum number of $X$-stabilizers and minimum number of $Z$-stabilizers~\cite{FujiiTokunaga2012}. Starting from the surface code in Fig.~\ref{fig:sc} we must therefore ask the question how to increase the number of $X$-stabilizers given the available actions in Fig.~\ref{fig:mv}. Since $X$-stabilizers are identified with vertices in the graph, repeatedly applying an action as displayed in Fig.~\ref{fig:mvb} will increase the number of $X$-stabilizers.
We hence expect a sequence of these actions to provide good strategies in this case, resulting in a surface code on a lattice with low connectivity.
Indeed, we can confirm this by looking at surface codes constructed by agents that have been successfully applied to this task. In Fig.~\ref{fig:graphiidZ}, a particularly interesting example solution is shown: It can be seen that the distance of the code w.r.t. $Z$-errors along one cycle has been increased from $3$ to $4$ by inserting $4$ new qubits at very specific points in the lattice. In other words, any logical $Z$-operator crossing the lattice from left to right in Fig.~\ref{fig:graphiidZ} is a product of at least $4$ single-qubit $Z$-operators. Just looking at the initial lattice in Fig.~\ref{fig:sc} it is not obvious that this can be done with only 4 actions. This is already a nontrivial result for this, at first glance, simple problem.
%%

% 9X,9Z
Now, let us consider the well-understood scenario of a i.i.d. error channel. That is, each qubit in the quantum computer is subject to a quantum channel of the form
\begin{align}
\mathcal{E}(\rho)=p_X\: X\rho X+p_Z Z\rho Z + (1-p_X-p_Z)\rho,\label{eq:errorXZ}
\end{align}
where we choose $p_X=p_Z=0.09$ such that either $X$- or $Z$-errors can happen with probability $0.09$ everywhere. Otherwise the task remains the same as before. We can see from Fig.~\ref{fig:resultiidXZ} that the agent again learns to improve the logical error rate while optimizing over the number of required qubits.

%check
The optimal surface code to protect against depolarizing noise is defined on a square (self-dual) lattice~\cite{FujiiTokunaga2012}. That is, the best known strategy to grow the surface code using the actions available (see Fig.~\ref{fig:mv}) is to retain the same connectivity both on the primal and dual lattice such that there are the same number of $X$- and $Z$-stabilizers. Indeed, looking at some examples from agents trained on this task, we can confirm that the agents also learn this strategy: The most successful agents end up creating surface codes where the number of $X$- and $Z$-stabilizers differ by at most~$1$. On average, the ratio between number of $X$- and $Z$-stabilizers is $1.06$ with standard deviation of $0.15$. The corresponding primal and dual lattices tend to have similar average connectivities, too. The average ratio between primal and dual connectivity is~$1.06$ with standard deviation of~$0.15$.

\subsection{QEC codes for correlated error channels}\label{sec:res_spatial}
%\subsection{Protecting against error channels with spatial dependencies}

%spatial dependencies
Next, let us tackle a complicated, practical scenario where the optimal strategy is unknown:
We consider spatial dependencies on top of an i.i.d. error channel.
This particular situation is motivated by the common problem of manufacturing defects. Correlated errors arising e.g., from manufacturing defects can be present in any fault-tolerant architecture because active error correction requires multi-qubit operations such as stabilizer measurements. Most architectures using topological codes have a spatial layout which can be associated with the lattice of the surface code~\cite{Barends-Martinis2014, Amini-Wineland2010} such that correlated errors will most likely be local.
Consider for instance an ion trap quantum computer using an arrangement of multi-zone Paul trap arrays~\cite{Bowler-Wineland2012}. Dephasing is the prevalent noise in ion traps~\cite{Schindler-Blatt2013}, so we would already have to consider an asymmetric error channel. Moreover, due to e.g., a manufacturing defect, two Paul traps in this arrangement could fail and produce $X$-errors on internal qubits.

\begin{figure}[ht!]
\begin{center}
	\vspace{0.5cm}
	\includegraphics[width=0.46\textwidth]{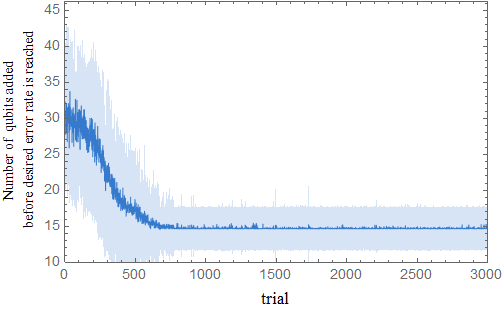}
	\caption{
		Learning curve for the task specified in Sec.~\ref{sec:res_spatial} averaged over $60$ agents. The shaded area shows the standard deviation. Here the quantum memory is subject to a noise channel with spatial dependencies as in Eq.~(\ref{eq:errorSpatial}). The initial logical error rate is as  high as $P_L^{\textrm{init}}\approx0.28$. Details about parameters are given in Appendix~\ref{app:param}.
	}\label{fig:resultSpatial}
\end{center}
\end{figure}

\begin{figure}[ht!]
\begin{center}
\includegraphics[width=0.4\textwidth]{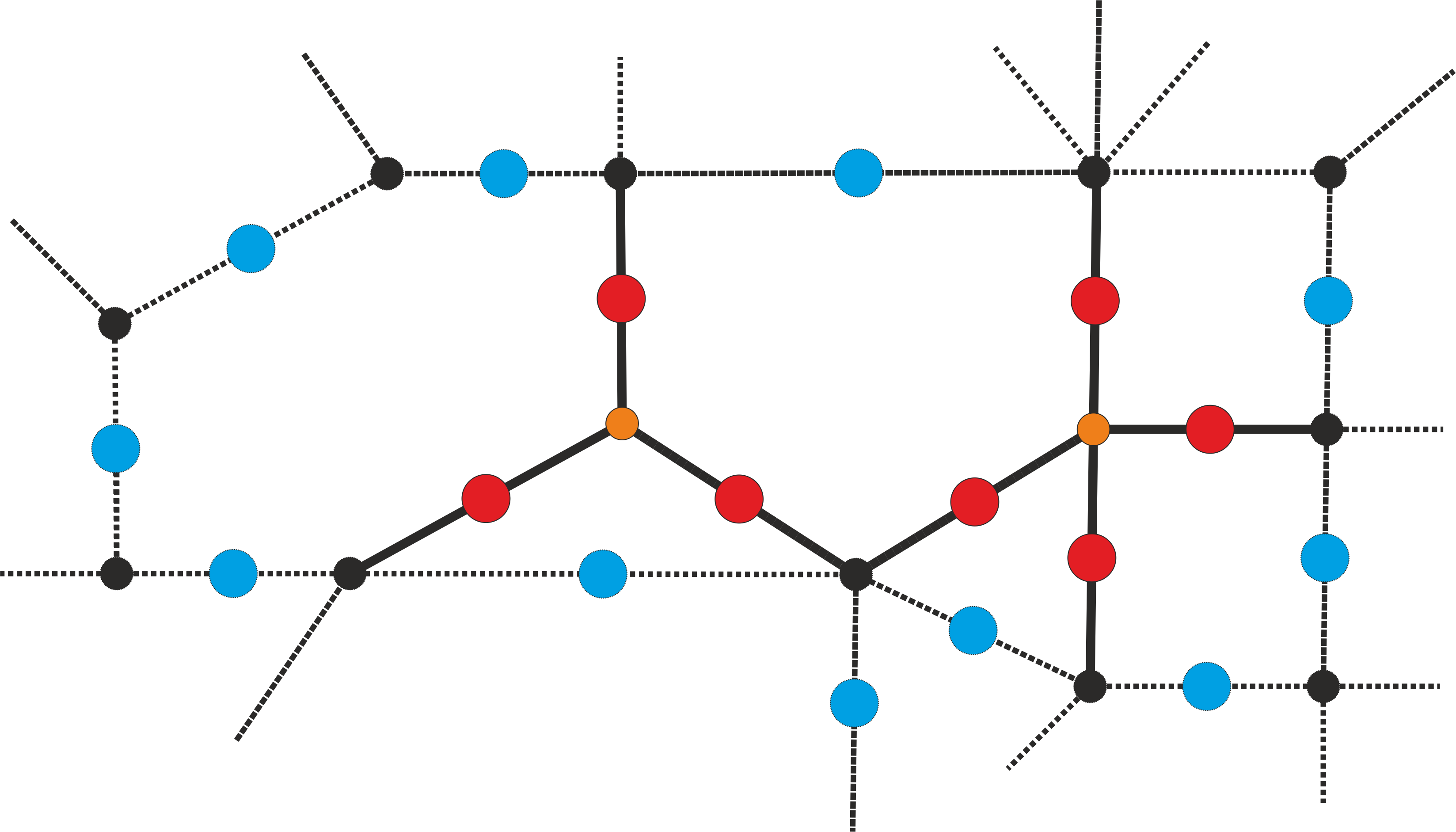}
	\vspace*{-3mm}
	\caption{%(Color online)
		Part of the final dual lattice constructed by one of the most successful agents in the task described in Sec.~\ref{sec:res_spatial} where noise has spatial dependencies (see Eq.~(\ref{eq:errorSpatial})). Qubits that are affected by high $X$-error probabilities are colored in red and surround flawed plaquettes colored in orange.
	}
	\label{fig:strategies}
\end{center}
\end{figure}

To be precise, consider the error channel in Eq.~(\ref{eq:errorXZ}), with $p_X=0.02$ and $p_Z=0.1$. This is similar to the simplest case in Sec.~\ref{sec:res_iid} where $p_X=0$.
In our construction, a correlated error as motivated above is modeled by an increased $X$-error rate on edges in the neighbourhood $\mathcal{N}(i)$ of a vertex or plaquette $i$. Here, we choose two neighbouring plaquettes $i,j$ and modify the error channel as follows,
 \begin{align}
 \mathcal{E}_k(\rho)=p_{X,k}\: X\rho X+p_{Z,k} Z\rho Z + (1-p_{X,k}-p_{Z,k})\rho,\label{eq:errorSpatial}
 \end{align}
where $k$ labels the qubits. We further assume that all qubits have base error rates of $p_{X,k}=0.02$ and $p_{Z,k}=0.1$. In addition, the base probability $p_{X,k}$ of an $X$-error on qubit $k$ is increased by $0.5$ if $k\in \mathcal{N}(i)\cup\mathcal{N}(j)$ and $p_{X,k}=1$ if $k\in \mathcal{N}(i)\cap\mathcal{N}(j)$.
This serves two purposes. First, we can evaluate the behavior of the agent with regard to the two %special
plaquettes $i,j$. How is the lattice structure changed in and around these plaquettes? Second, we can understand how the agent handles a deterministic error on the edge neighbouring both $i,j$. Will this edge be treated differently? In fact, it is far from clear what the optimal strategy is given the available actions displayed in Fig.~\ref{fig:mv}. Nevertheless, we observe that the PS agent learns to protect the surface code against this error channel while optimizing the used resources, see Fig.~\ref{fig:resultSpatial}.
Now, let us evaluate successful strategies employed by the best agents in this scenario. In Fig.~\ref{fig:strategies} the relevant part of the final dual lattice as constructed by a successful agent is depicted. We can see that the agent increases the number of $Z$-stabilizers in the immediate vicinity of the flawed plaquettes, decreasing the connectivity of affected plaquettes. Interestingly, the agent also finds a way of moving the flawed plaquettes apart given the available actions, thereby removing any deterministic error appearing on the edge between these plaquettes. At the same time, due to the prevalent $Z$-errors, connectivity throughout the lattice is balanced between vertices and plaquettes: the average connectivity of the dual lattice is $3.9$ and $4.1$ for the primal lattice. Similarly, the ratio between $X$- and $Z$-stabilizer is $14/15$.

\subsection{QEC codes with changing requirements}\label{sec:res_rewardChange}
\begin{figure}[ht!]
\begin{center}
	\vspace{0.5cm}
	\includegraphics[width=0.48\textwidth]{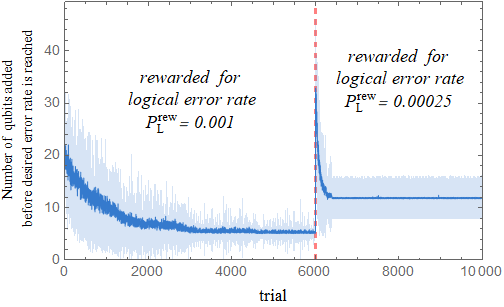}
	\caption{
		Learning curve for the task specified in Sec.~\ref{sec:res_rewardChange} averaged over $60$ agents. The shaded area shows the standard deviation.
		In this task the agent is tasked to reduce the logical error rate below $P_\mathrm{L}^{\mathrm{rew}}=0.001$ while the code is subject to the error channel in Eq.~(\ref{eq:errorZ}) with $p=0.1$. After 6,000 trials the same agent is tasked to reduce the error rate even further below $P_\mathrm{L}^{\mathrm{rew}}=0.00025$. Details about parameters are given in Appendix~\ref{app:param}.
	}\label{fig:rewQuarter}
\end{center}
\vspace*{-3mm}
\end{figure}

%%
%changing reward
So far, we have kept the rewarded logical error rate fixed. However, it is in general also desirable to be able to adapt to stricter target thresholds if required. %We therefore also wish to test if transfer learning is useful in this scenario.
We therefore consider a scenario where the error channel is initially fixed to that of Eq.~(\ref{eq:errorZ}) with $p=0.1$. Then, after having learned the optimal strategy from before, the agent is tasked with further decreasing the logical error rate to a quarter of the initial value. As one can observe from Fig.~\ref{fig:rewQuarter}, the agent can indeed draw on its knowledge from the first stage of the task to find a solution for the second stage.

In the particular scenario of varying target thresholds, the performance of the agent can potentially be improved by issuing a reward which is proportional to the inverse logical error rate. This provides more feedback about the reward landscape in form of a gradient which can be followed. Such a modified reward landscape can also be exploited by simpler optimization algorithms such as Simulated Annealing~\cite{KirkpatrickGelattVecchi1983}. In addition, there exist other approaches for optimizing QEC procedures for fixed error models~\cite{ReimpellWerner2005, KosutLidar2009} and for variational optimization based on directly available experimental data (QVECTOR~\cite{JohnsonEtAl2017}), which have been shown to improve noise robustness for small numbers of qubits.
However, our main focus here lies on exploring the observed ability to capitalize on previously obtained knowledge encoded in a memory. As we will see in Sec.~\ref{sec:on-off-line}, this ability can be viewed as part of a more general transfer learning~\cite{Thrun1996,WeissKhoshgoftaarWang2016} skill that represents a main strength of RL for QEC. While common optimization methods usually do not feature a memory, it should be noted that RL can be combined with optimization techniques to improve initial exploration, thereby ameliorating the training data used for learning~\cite{anonymous2020improving, Levine:2013:GPS:3042817.3042937}.

\section{Simulation vs. Experiment}\label{sec:on-off-line}

%on-line optimization in experiments
In the previous sections, the main focus has been to determine whether RL agents in our framework are able to adapt and optimize a surface code quantum memory to various types of noise with different requirements.
We have found that this is indeed the case, showcasing the prospect of RL in on-line optimization of quantum memories. Although we have evaluated our procedures only via simulations, our results suggest that such approaches can be successful also in practice in future laboratory experiments.
This is because our framework for optimizing QEC codes in Fig.~\ref{fig:agent-environment} is independent of whether the environment is a real or simulated quantum device.
In either case, the interface between environment and agent remains unchanged.
For instance, we estimate the logical error rate of the quantum memory using a Monte Carlo simulation. In a real device, this can be done by preparing a quantum state, actively error correcting for a fixed number of cycles, and then measuring the logical operators by measuring all data qubits in a fixed basis. The logical error rate should then be interpreted as the probability of a logical error per QEC cycle.
Repeating this measurement provides an estimation of the lifetime of the quantum memory. Moreover, the code deformations which constitute the actions available to an agent are designed with a physical device in mind~\cite{BravyiKitaev1998, BombinMartinDelgado2009}.

\begin{figure*}[t!]
	\centering
    %%%trim={<left> <lower> <right> <upper>}
	\subfigure[]{\includegraphics[width=0.3\textwidth,trim={1cm 0mm 1cm 0mm}]{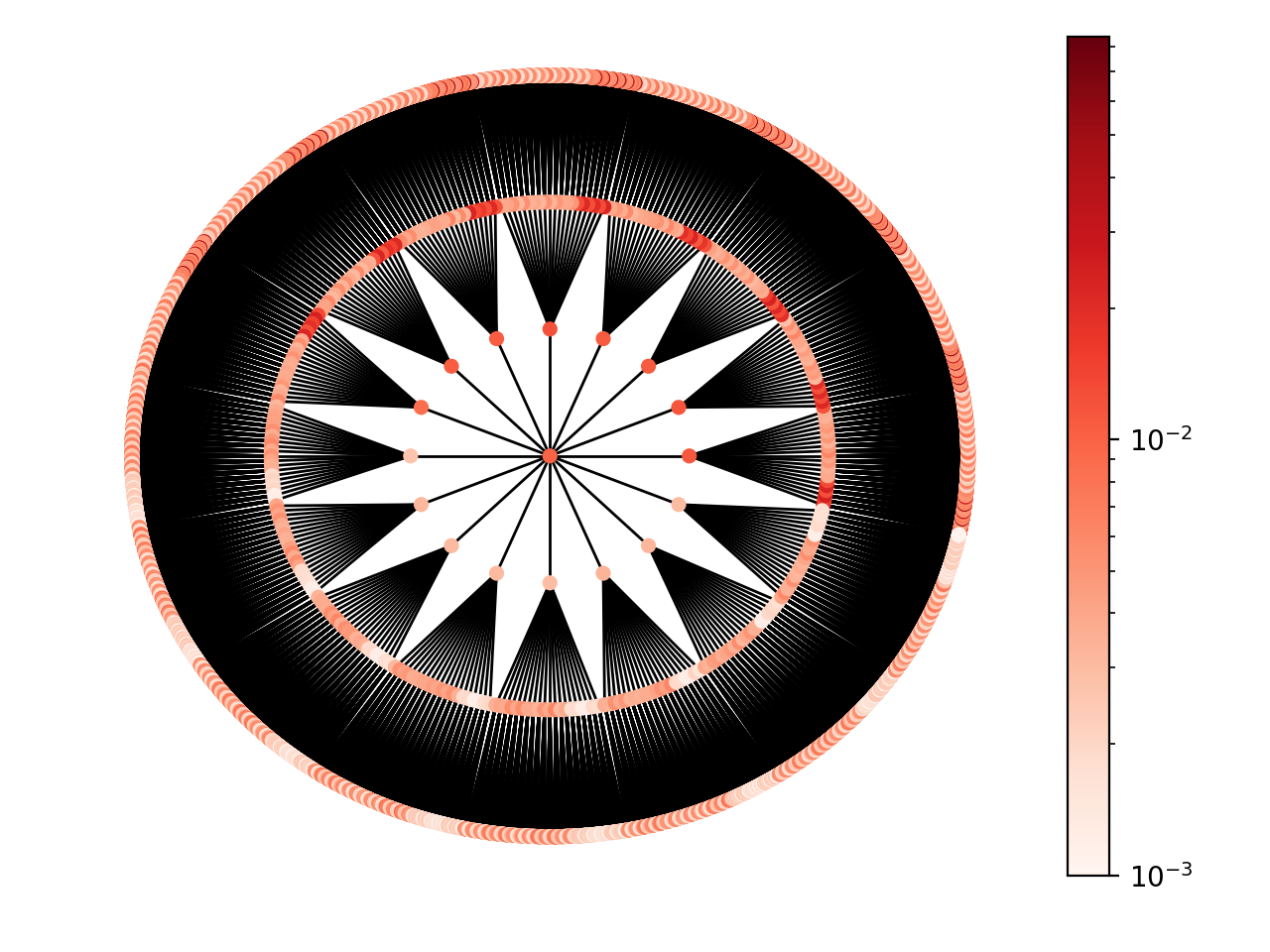}}\hspace{0.30cm}
	\subfigure[]{\includegraphics[width=0.3\textwidth,trim={1cm 0mm 1.25cm 0mm}]{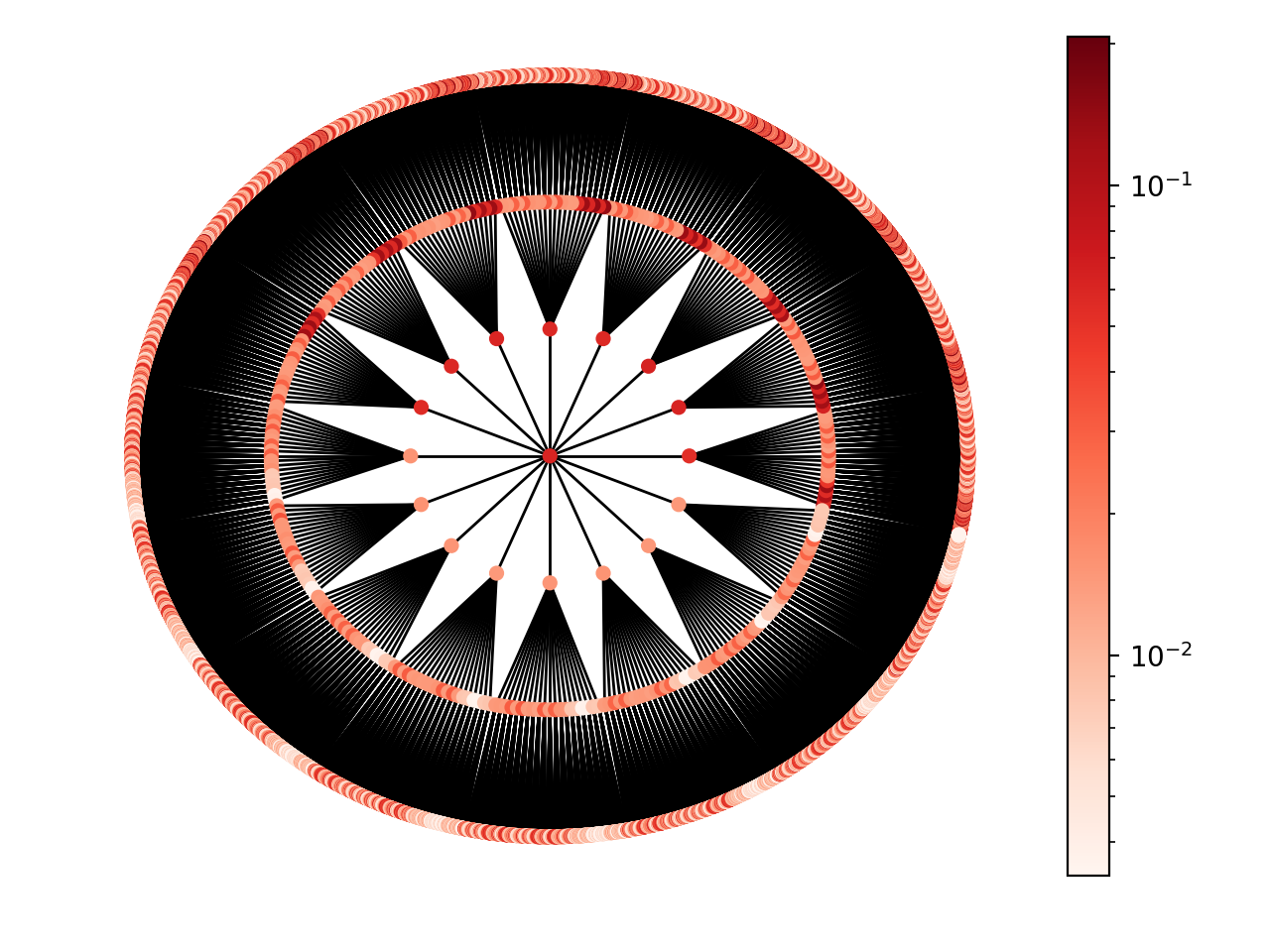}}\hspace{0.30cm}
    \subfigure[]{\includegraphics[width=0.33\textwidth,trim={0cm 0mm 0.75cm 0mm}]{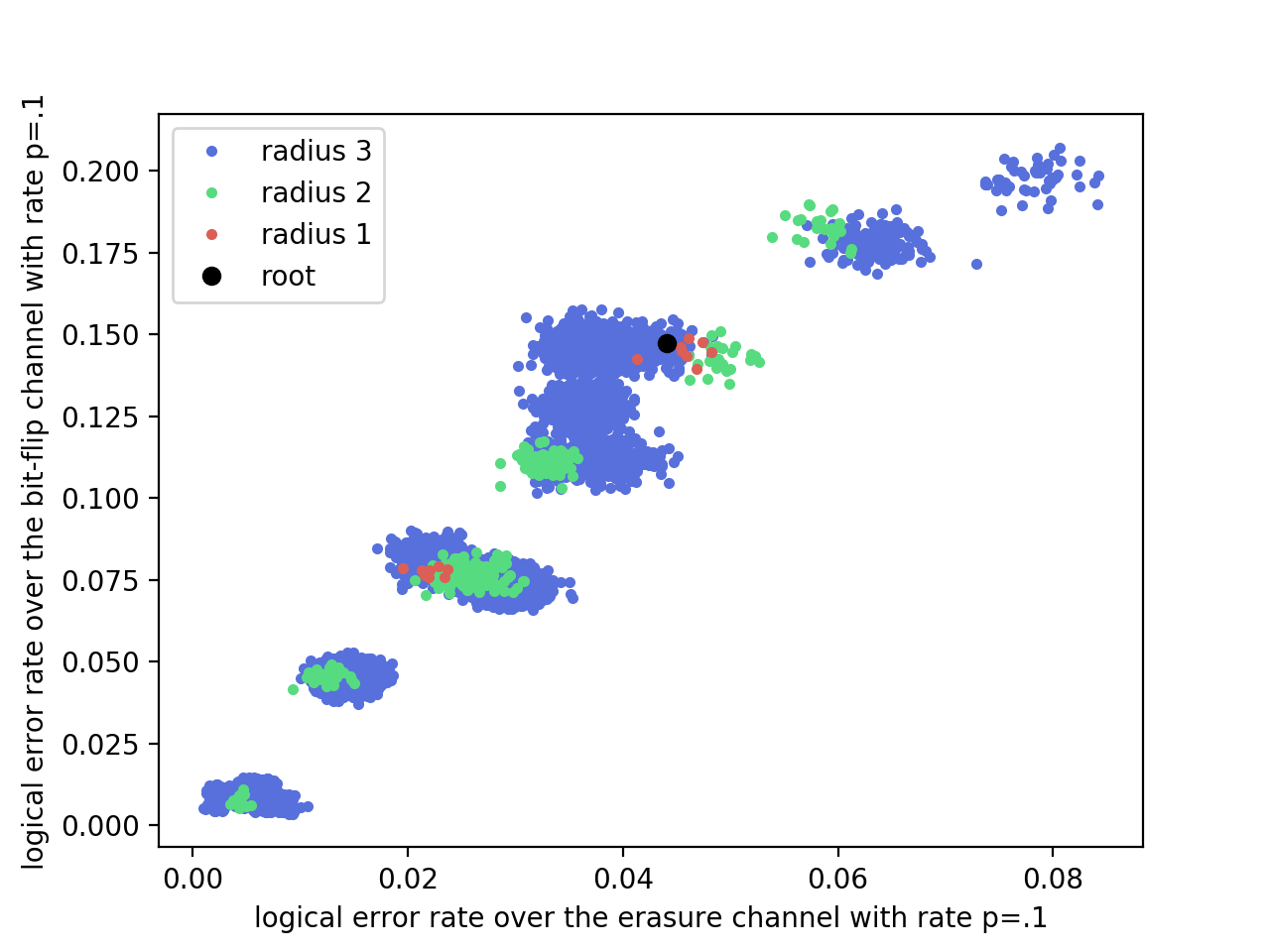}}	
\vspace*{-4mm}
	\caption{Visualization of the search space and comparison of logical error rates estimated by (a) SQUAB~\cite{DelfosseZemor2017} (using an approximated Pauli error model) and (b) the Union Find decoder~\cite{DelfosseNickerson2017} (using a standard Pauli error model). Each node in the tree corresponds to an individual surface code on a torus and edges imply that two codes are connected by an action as in Fig.~\ref{fig:mv}.
		The central node is the $2\times 2$ surface code on a square lattice. We explore all codes up to a distance $3$ from the root. The color shading indicates the logical error rate.
		(a) The logical error rate is estimated using SQUAB with 10,000 trials and erasure rates $p^Z_e = 0.1$, $p^X_e=0$.
		(b) The logical error rate is estimated using the Union-Find decoder with 10,000 trials and a $Z$-error rate $p_Z = 0.1$. While the logical error rates differ in both cases, the behavior
		is qualitatively the same for a code subject to an asymmetric erasure channel and the same code affected by Pauli noise.
		That is, codes found to provide lower logical error rates w.r.t. other codes for one noise model also provide lower logical error rates than other codes for the other error channel.
        (c) The scatter plot compares the logical error rates of the erasure channel (horizontal axis) and bit-flip channel (vertical axis) with identical rates $p_e^Z=p_Z=0.1$ for the same codes, where each dot corresponds to a different code. The observed alignment of clusters on the diagonal can be seen as a qualitative sign for the agreement of the error models.
		%That is, the difference in logical error rates as estimated by SQUAB $P_\mathrm{L}^{\mathrm{SQ}}$ and the Union Find decoder $P_\mathrm{L}^{\mathrm{UF}}$ is just a matter of magnitute: for almost any code $C$ in the space of all surface codes, $P_\mathrm{L}^{\mathrm{SQ}}(C)= \alpha P_\mathrm{L}^{\mathrm{UF}}(C)$ with the same constant $\alpha$.
	}
	\label{fig:squabVSunionfind}
\end{figure*}
%%

%from simulation to experiment
Ultimately, one is of course interested in performing optimization on actual quantum devices. However, as long as these devices are sufficiently small-scale to allow for simulations that are (much) faster than the timescale of operating the actual device, it is advantageous to perform the bulk of optimizations off-line in simulations before transferring the results to the actual device for further optimizations. In other words, to fully capitalize on the important features of transfer-learning and pre-training in quantum-applied machine learning, it is crucial to ensure that the simulations are as efficient as possible, and that the chosen RL model is able to transfer experience from simulations to real environments. In the following, we therefore discuss the efficiency and practicality of our simulations and show that the RL agents we consider are capable of transfer learning.

\subsection{Simulation Efficiency of QEC}\label{sec:res_efficiency}
%fast simulations
%error model + decoder

To provide sufficiently efficient off-line simulation, the individual components of our QEC simulation have been carefully selected. For instance, note that stabilizer QEC can be simulated efficiently classically~\cite{CleveGottesman1997, AaronsonGottesman2004, DiVincenzoShor1996, AndersBriegel2006}. However, estimating the logical error rate, which requires a large number of samples from QEC simulations, remains computationally expensive. In our simulations, we hence make use of a simplified error model to allow for faster estimations of logical error rates. The simplified error model is based on the quantum erasure channel~\cite{BennettDiVincenzoSmolin1997} since there exists a linear-time maximum likelihood decoder for surface codes over this channel which is optimal both in performance and speed~\cite{DelfosseZemor2017}.
The use of this software, SQUAB~\cite{DelfosseIyerPoulin2016, SQUABweb}, within our RL framework is hence providing the means for sufficiently fast exploration of the space of surface codes.
In essence, the erasure error channel is very similar to a Pauli error channel where the location of an error is known exactly. More specifically, we introduce an asymmetric erasure channel where we choose two erasure probabilities $p^X_e, p^Z_e$ that can have spatial and temporal dependencies. Since $X$ and $Z$ stabilizers are separated in the surface code, error correction of $X$ and $Z$ errors can also be treated independently. In the simulation of $X$ errors over the erasure channel, we erase each qubit $\rho$ in the surface code with probability $p_e^X$, and replace it by a mixed state of the form $\rho'= \frac{1}{2}(\rho+ X\rho X)$. The set of erased qubits is known. The simulation proceeds analogously for $Z$ errors.

%erasure channel: practicality
Since our simulations cover setups with up to $68$ qubits, we have indeed good reason to believe that our simulations are sufficiently efficient to optimize QEC codes for near-term quantum devices without the need of experimental data.
However, one may argue that our software only provides a simulation for a simplified noise model, the erasure channel.
In order to prove that the results are relevant in practice, we compare logical error rates of a set of small surface codes subject to erasure errors to the rates obtained from simulating standard Pauli noise on the same codes.
To this end, we assume that each qubit is affected independently by a $Z$-type Pauli error and we perform error correction with both SQUAB, and the Union-Find decoder introduced in Ref.~\cite{DelfosseNickerson2017}.
We report the average logical error rate for each code after 10,000 trials with an erasure rate $p^Z_e=0.1$ (SQUAB) and $Z$-error
rate $p_Z = 0.1$ (Union Find).
In Fig.~\ref{fig:squabVSunionfind} we observe qualitatively the same behavior:
codes that are considered to perform well according to the estimation using SQUAB are also considered to perform well using the Union-Find decoder.
The difference between the two error channels lies predominantly in the magnitude of the logical error rates.
It is therefore better to select codes using SQUAB, allowing for a faster and thus more precise exploration of the space of topological codes.

\subsection{Transfer learning in QEC}\label{sec:res_tf}

%transfer learning
The usefulness of off-line simulations for QEC code optimization emerges from the application of the results to on-line settings. In this transition, deviations of the error model used in the simulation from the actual noise, or dynamical changes of the latter can lead to non-optimal performance if no further action is taken.  Here, machine learning comes into play. That is, a central agenda of machine learning is to develop learning models that are capable of successfully applying previously obtained knowledge in different environments. This possibility, called transfer learning~\cite{Thrun1996}, is an active area of research in general AI~\cite{WeissKhoshgoftaarWang2016}.
Here we should note that the ability to \textit{transfer} knowledge is indeed much desired but not manifestly present in all machine learning models. At the same time, there is the risk of confusion with the more generally encountered ability for generalization~\cite{SaittaZucker2013, MelnikovMakmalDunjkoBriegel2015}. Let us quickly illustrate the difference: On the one hand, generalization is the ability of a learning agent to be effective across a range of inputs and knowing what to do in similar situations. Transfer learning, on the other hand, is the ability to attain knowledge in one scenario, and then being able to use this knowledge in a different (new but related) learning setting. One of the objectives of transfer learning is to jump start the learning process in a new, but similar scenario, and is fundamentally linked to the AI problem of \textit{learning how to learn}~\cite{PatriciaCaputo2014, Thrun1996}.

Transfer learning can provide a significant improvement in performance as compared to optimization approaches without learning mechanisms, as well as considerable reductions of learning times in comparison to untrained RL agents. Put simply, agents capable of transfer learning do not have to start from scratch every time the environment or task changes slightly~\cite{TommasiCaputo2009,TommasiOrabonaCaputo2010,AytarZisserman2011, MelnikovPoulsenNautrupKrennDunjkoTierschZeilingerBriegel2017}. As we will discuss in this section, within the RL setting we consider here, agents trained on the original simulations may \emph{transfer} their experience from simulations to practical applications. The usefulness of this approach will of course depend on how similar the simulated error model is to the real error channel.

\begin{figure*}[t!]
\begin{center}
	\subfigure[]{\includegraphics[width=0.48\textwidth]{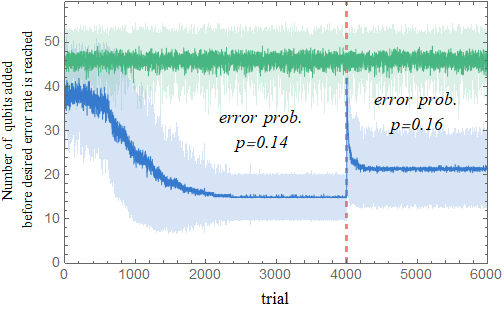}
		\label{fig:comparison}}\hspace{0.30cm}
	\subfigure[]{\includegraphics[width=0.44\textwidth]{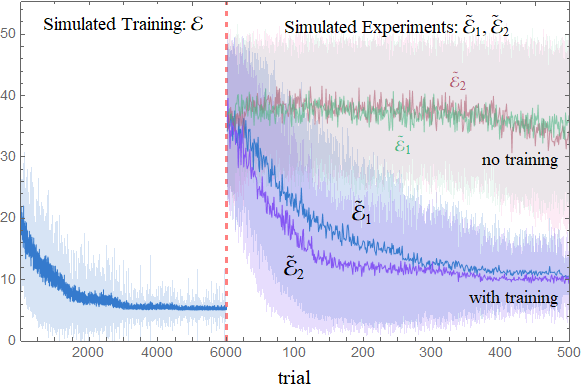}
		\label{fig:TL}}
	\vspace*{-3mm}
	\caption{%(Color online)
		This figure shows the results for two tasks as considered in Sec.~\ref{sec:res_tf} which explore the potential of transfer learning. The agent is tasked to optimize a QEC code w.r.t. its resource efficiency. Resource efficiency is reflected in the number of qubits added to the code before a desired logical error rate $P_\mathrm{L}^{\mathrm{rew}}$ is reached. The shaded area shows the standard deviation for $60$ agents. (a) At first, the code is subject to an error channel as in Eq.~(\ref{eq:errorZ}) with $p=0.14$ and $P_\mathrm{L}^{\mathrm{rew}}=0.001$. The initial logical error rate is $P_L^{\textrm{init}}\approx0.019$. At trial $4,000$ the error probability is increased to $p=0.16$.  The blue curve depicts the results in this scenario averaged over 60 agents. As comparison, the green line is the average over 40 agents that are directly tasked to protect against an error channel with $p=0.16$ without having been trained on the simpler noise model. These agents encounter an initial logical error rate of $P_L^{\textrm{init}}\approx0.028$. (b) A transfer learning scenario where $60$ agents (blue curve) are trained on a simple, unrealistic error model $\mathcal{E}$ [see Eq.~(\ref{eq:errorZ})] before being subjected to either one of the more realistic noise channels $\tilde{\mathcal{E}}_1, \tilde{\mathcal{E}}_2$ described in Eq.~(\ref{eq:errorTL}) and~(\ref{eq:errorTL2}), respectively. For the second stage we have randomly selected one of the most successful agents from the first stage. As comparison, the green and pale rose curve depict an average over $60$ agents with the same parameters, subjected to the channels  $\tilde{\mathcal{E}}_1, \tilde{\mathcal{E}}_2$, respectively, but without pre-training. Details about parameters are given in Appendix~\ref{app:param}.
	}
	\label{fig:resultTL}
\end{center}
\end{figure*}

In order to demonstrate the potential of transfer learning in the context of optimizing and adapting quantum memories, we consider a scenario where the agent is first trained on one error model $\mathcal{E}$, and then encounters a different error model $\tilde{\mathcal{E}}$.
This change in noise could occur not only because the agents are switched from a simulated training scenario to an actual experiment but also, e.g., due to environmental changes, equipment malfunctions or malicious attacks.
Generally, under the assumption that the noise can only vary slightly, we can expect learning to be beneficial since strategies that helped in one scenario should still be favorable in protecting against similar error channels. Then, an RL algorithm can further optimize over a given policy and re-adjust to the new environment.
This scenario capitalizes on the strength of RL since it requires a long-term memory that can be exploited between tasks. If exploration of the search space is encouraged over exploitation of rewards, the agent's memory contains global information about the environment. In other words, given some sub-optimal QEC code, such an agent knows a strategy to adapt this code efficiently such that the desired logical error rate is reached.
In fact, the feedback which is given only if a certain logical error rate has been achieved, is specifically designed to encourage exploration of the search space.

The specific choice of error models for this scenario is partially motivated by the results in Fig.~\ref{fig:squabVSunionfind}. There we observe that the main difference between the simulated erasure and more realistic Pauli error channels lies in the magnitude of the estimated logical error rate.
We therefore now consider a quantum memory which is first subjected to the simple noise channel in Eq.~(\ref{eq:errorZ}) with $p=0.14$ and the agent is tasked again to reduce the logical error rate below $P_\mathrm{L}^{\mathrm{rew}}=0.001$. Then, after having learned a beneficial strategy for finding a good QEC code, the same agent is confronted with an increased error rate of $p=0.16$.
Fig.~\ref{fig:comparison} shows that, in this second stage of the task, the agent benefits from having learned to protect the quantum memory against the first error channel.
In particular, we see from Fig.~\ref{fig:comparison} that agents not initially trained on the first noise channel behave randomly and do not find reasonably good strategies to deal with the high noise level. This is because at a physical error rate of $p=0.16$ the initial code requires many modifications before the desired logical error rate can be reached. In fact, the required number of basic modifications is so large that a random search is just not sufficient to find a reasonably good code in the allotted time of $6,000$ trials. Although the observed behaviour of the learning agents showcases the benefit of a memory, the error models are nevertheless too similar to reveal the potential for \emph{transfer learning}.

Let us therefore consider another scenario with more drastic changes in the noise model. In particular, we attempt to model a setting more closely resembling the transfer of knowledge from simulation to experiment: That is, we start by training $60$ agents on an error channel $\mathcal{E}$ that captures partial knowledge of the real error channel disturbing a quantum system. Since dephasing is the prevalent noise in many quantum computation architectures~\cite{Schindler-Blatt2013,AliferisBritoDiVincenzoPreskillSteffenTerhal2009,ShulmanDialHarveyBluhmUmanskyYacoby2012}, we choose the error model from Eq.~(\ref{eq:errorZ}) with $p=0.1$ for training. Here, we deliberately neglect generally more realistic Pauli errors to showcase the transfer learning ability, but we allow sufficiently many iterations so the agents can optimize their behavior w.r.t. to the unrealistic noise model. Then, in the second stage of this scenario, we select (at random) one of the best agents and confront it with a more realistic error model $\tilde{\mathcal{E}}_1$ which features both $X$-, and $Z$-errors as well as spatial correlations. The error model, similar to that in Eq.~(\ref{eq:errorSpatial}), is
\begin{align}\label{eq:errorTL}
\tilde{\mathcal{E}}_{1,k}(\rho)=p_{X,k}\: X\rho X+p_{Z,k} Z\rho Z + (1-p_{X,k}-p_{Z,k})\rho,
\end{align}
where $k$ labels the qubit the channel acts on. All qubits have base error rates of $p_{X,k}=0.02$ and $p_{Z,k}=0.14$. In addition, the base probability $p_{X,k}$ of an $X$-error on qubit $k$ is increased by $0.15$ if $k\in \mathcal{N}(i)$ where $i$ labels one specific plaquette of the lattice to model, e.g., a manufacturing defect as in Sec.~\ref{sec:res_spatial}.

To further challenge the learning algorithm in this scenario, we only allow a limited number of trials in the second stage (less than 10\% of the first stage) since any actual experiment would be much more expensive than simulations. Despite the significant difference between the error models $\mathcal{E}$ and $\tilde{\mathcal{E}}_1$, we observe in Fig.~\ref{fig:TL} that the agents are able to significantly capitalize on the knowledge obtained from the initial training simulations (blue curve). In contrast, the same agent without pre-training (plotted in green in Fig.~\ref{fig:TL}) barely learns at all during the allotted number of trials.

This advantage is indeed remarkable. However, it would be of no practical use if the benefits from transfer learning were not \textit{robust} to variations in the experimental noise channel. That is, we expect the advantage of transfer learning to extend to other noise channels, too. Therefore, let us exchange the error channel $\tilde{\mathcal{E}}_1$ by another, new model $\tilde{\mathcal{E}}_2$ where we replace spatial correlations by a doubled base $X$-error rate, i.e.
\begin{align}\label{eq:errorTL2}
\tilde{\mathcal{E}}_2(\rho)=p_X\: X\rho X+p_Z Z\rho Z + (1-p_X-p_Z)\rho,
\end{align}
with $p_Z=0.14$ and $p_X=0.04$. Now, choosing the same pre-trained agent as before and transferring it to the new setting, we find that, despite the drastic difference between $\tilde{\mathcal{E}}_1$ and $\tilde{\mathcal{E}}_2$, the advantage gained from transfer learning is still substantial (which can be concluded from the comparison between the purple and pale rose curve in Fig.~\ref{fig:TL}).
This remarkable and robust advantage showcases the benefits of transfer learning for QEC in resource-limited, near-term quantum devices.

\section{Discussion}\label{sec:discussion}

Reinforcement learning (RL) has recently seen a great deal of success, from human-level performance in Atari games~\cite{Mnih-Hassabis2015} to beating the world champion in the game of Go~\cite{Silver-Hassabis2017}. In 2017, RL was included in the list of 10 breakthrough technologies of the year in the MIT technology review~\cite{MITTechReviewWeb}. As machine learning is claiming its place in the physicist's toolbox~\cite{Zdeborova2017}, RL is starting to appear in quantum physics research~\cite{MelnikovPoulsenNautrupKrennDunjkoTierschZeilingerBriegel2017, ItenMetgerWilmingDelRioRenner2019, SwekeKesselringVanNieuwenburgEisert2018, FoeselTighineanuWeissMarquardt2018, AndreassonJohanssonLiljestrandGranath2019, AugustHernandez-Lobato2018}.

In this work, we have presented an RL framework for adapting and optimizing quantum error correction (QEC) codes. This framework is based on an RL agent that interacts with a quantum memory (and its classical control), providing the latter with instructions for modifications of the code to lower the logical error rate below a desired threshold. For the simulations discussed here, the quantum memory is realized as a surface code to which the agent may add qubits by way of fault-tolerant code deformations~\cite{BravyiKitaev1998,BombinMartinDelgado2009}. The agent receives a reward once the specified logical error rate is reached. The internal structure of the agent has been modeled within the Projective Simulation~\cite{BriegelDeLasCuevas2012} approach to RL.
%which has recently been successfully employed for standard RL problems~\cite{MautnerMakmalManzanoTierschBriegel2015, MelnikovMakmalBriegel2018, MelnikovMakmalDunjkoBriegel2015}, advanced robotics applications~\cite{HanglUgurSzedmakPiater2016}, and even to design new quantum experiments~\cite{MelnikovPoulsenNautrupKrennDunjkoTierschZeilingerBriegel2017}.
The classical control system estimates the logical error rate in the simulated QEC procedure using an optimal linear-time decoder~\cite{DelfosseIyerPoulin2016, SQUABweb}.

Our results demonstrate that the agents learn to protect surface code quantum memories from various types of noise, including simple i.i.d. errors, but also more complicated correlated and non-isotropically distributed errors. In particular, this RL approach provides interesting solutions for nontrivial QEC code optimization problems. A particularly noteworthy feature is the ability of the agents to transfer their experience from one noise model or task to another even if they are seemingly very different. This suggests that such a QEC strategy based on RL can be used to switch from off-line optimization to on-line adaptive error correction. That is, provided a reasonable guess for the type of expected errors, one may start by training an RL agent on simulations. Then, the trained agent can be used to bootstrap optimization in the actual hardware.

The scope of our simulations has been designed specifically with such applications to current state-of-the-art quantum computing platforms~\cite{FriisMartyEtal2018,Zhang-Monroe2017,Bernien-Lukin2017} in mind. Starting with 18 initial qubits, the agents we consider are able to extend this number by up to 50 additional qubits, thus also accounting for expected near-term technology developments. A potential bottleneck for extensions to much larger qubit numbers lies in the scaling behavior of the learning complexity. There, one expects that the increase in learning time scales unfavorably with the increase in the size of the percept space (and action space, both of which depend on the number of qubits). We envisage that this problem can be circumvented through parallel processing by assigning individual agents to different surface code patches of a fixed size. All agents can then operate in parallel, allowing one to exploit a number of already available RL techniques for parallel learning that explore the use of shared experience between agents~\cite{Kretchmar2002,BarrettDugganHowley2014}. Due to the local structure of the problem of optimizing topological QEC codes, parallel RL may potentially yield close to optimal policies even for significantly larger lattice sizes. Moreover, note that the search of our RL agent to find QEC codes is largely specified by the history of performed actions. Thus, one might be able to reduce the complexity of RL approaches by employing long short-term memories (LSTMs)~\cite{HochreiterSchmidhuber1997}, in order to enable deliberation with respect to a few past actions. Since actions are also localized, this might yield enough information to evaluate an optimal policy that is partially blind to the lattice size.

At the same time, the practical success of machine learning techniques also depends on suitably narrowing down the optimization problem. For example, in Ref.~\cite{FoeselTighineanuWeissMarquardt2018}, neural networks were used to determine sequences of quantum gates and measurements to protect a logical qubit. This allows the algorithm to search the whole space of quantum circuits. However, this comes at a cost. The space of all possible QEC strategies that the algorithm explores is so vast that scaling inevitably becomes an issue. While this work demonstrates a successful strategy on up to $4$ data qubits subject to uncorrelated bit-flip errors, significant advances would be needed to generalize this approach to larger, potentially varying numbers of data qubits and more realistic noise. In contrast to the approach of~\cite{FoeselTighineanuWeissMarquardt2018}, the method developed here can adapt and optimize QEC codes in terms of their resource efficiency, i.e., the number of data qubits needed to achieve a desired maximal logical error rate, and operates without detailed information about or precise simulation of the underlying quantum states.

Moreover, let us note that the presented general framework for RL-based optimization of error correction codes goes beyond the specific simulations we have performed here. That is, the approach is directly applicable also if one wishes to consider RL paradigms other than PS, QEC codes other than surface codes, or noise models other than those considered. A particular contributor to this flexibility is that both the errors and the error decoding appear as part of a black-box environment to the agent, who only ever perceives whether the logical error rate is above or below target threshold. For instance, one is left with the option of choosing different decoders, and even of incorporating machine learning into the decoder itself~\cite{TorlaiMelko2017, BaireutherOBrienTarasinskiBeenakker2018, ChamberlandRonagh2018, SwekeKesselringVanNieuwenburgEisert2018, BaireutherCaioCrigerBeenakkerOBrien2018, Ni2018, MaskaraKubicaOConnor2018, LiuPoulin2018, AndreassonJohanssonLiljestrandGranath2019}. Having this freedom in choosing the decoder is particularly relevant if different types of QEC codes are considered, e.g., if one allows fault-tolerant code switching through code deformation~\cite{BombinMartinDelgado2009} or lattice surgery~\cite{PoulsenNautrupFriisBriegel2016} as part of the optimization. In summary, while the simulations have been carried out within a specific setting, this framework lays the groundwork for applying sophisticated RL algorithms to more general (topological) QEC codes and more realistic noise models~\cite{AharonovKitaevPreskill2006,NgPreskill2009,FowlerMartinis2014,NickersonBrown2019}.

\begin{acknowledgments}
	\vspace*{-2mm}
	HPN and HJB achnowledge support from the  Austrian Science Fund (FWF) through the projects DK-ALM: W1259-N27 and SFB BeyondC: F7102.
	HJB was also supported by the Ministerium f\"ur Wissenschaft, Forschung, und Kunst Baden-W\"urttemberg (AZ:33-7533.-30-10/41/1).
	NF acknowledges support from the Austrian Science Fund (FWF) through the project P 31339-N27, the START project Y879-N27, and the joint Czech-Austrian project MultiQUEST (I 3053-N27 and GF17-33780L).
	HPN would like to acknowledge helpful comments from Davide Orsucci, Robert Raussendorf and Lea Trenkwalder.
	ND would like to thank Michael Beverland, Krysta Svore, Dave Wecker and Nathan Wiebe for useful discussions.
\end{acknowledgments}

%\vspace{10cm}

\appendix
\section{Learning in Projective Simulation}\label{app:PS}
In Sec.~\ref{sec:rl_PS} we have given a brief introduction the projective simulation (PS) model for reinforcement learning (RL). Here we complete the description with a detailed account of how the $h$-values that govern the transition probabilities in Eq.~(\ref{eq:transitionprob}) are updated. In other words, we explain how learning manifests itself in the PS model. The update rule from time step $t$ to $t+1$ is
\begin{align}
h^{(t+1)}=h^{(t)}+\lambda^{(t)}g^{(t)}+\gamma(1-h^{(t)})\label{eq:update}
\end{align}
where the $g$-matrix, or so-called glow matrix, redistributes rewards $\lambda$ to past experiences such that experiences that lie further in the past are rewarded only by a decreasing fraction of $\lambda^{(t)}$. In other words, past experiences are remembered (they `\textit{glow}') less strongly and therefore less likely to influence future behaviour, which is  represented by a decreasing factor, i.e. the glow value. The glow matrix contains the long-term memory of the agent. A long-term memory is crucial when rewards are delayed, i.e., not every action results into a reward. The glow matrix is updated in parallel with the $h$-matrix. At the beginning, $g$ is initialized as an all-zero matrix. Every time an edge $(i,j)$ is traversed during the decision-making process, the associated glow value $g_{ij}$ is set to $\frac{M_i}{M_0}$ where $M_i$ is the number of actions available for percept $i$ and $M_0$ is the number of initial actions. In order to internalize how much of the reward $\lambda$ is issued to past experiences in Eq.~(\ref{eq:update}) the $g$-matrix is also updated after each interaction with the environment. Therefore, we introduce the so-called glow parameter $\eta\in[0,1]$ of the PS model and define an update rule as follows,
\begin{align}
g^{(t+1)}=(1-\eta)g^{(t)}.
\end{align}
Besides glow, the agent is also subject to a forgetting mechanism, which is presented by the parameter $\gamma$ in Eq.~(\ref{eq:update}). Effectively, the $h$-values are decreased by $\gamma\in[0,1]$ (but never below $1$) whenever an update is initialized. Here, the forgetting mechanism is used specifically to reinforce exploratory behavior. In order to save memory, we also introduce a deletion mechanism where unused percept clips are deleted, i.e., if the average of outgoing $h$-values is below a value $1+\delta$. However, deletion can only be applied to a specific clip once a certain number of rewarded interactions $\tau$ with the environment have passed since its creation. This parameter $\tau$ is hence called immunity time. Moreover, after any given trial, all clips that have been created during said trial are deleted if no reward was given at all.
Note that finding the optimal values for $\beta,\eta,\gamma,\tau$ and $\delta$ is generally a hard problem. However, it can be learned by the PS model itself~\cite{MakmalMelnikovDunjkoBriegel2016} at the expense of longer learning times. However, hyperparametrization is often easier and faster in PS than in comparable RL models~\cite{MelnikovMakmalBriegel2018}.

\section{Details of the environment}\label{app:env}
%classical control
In Sec.~\ref{sec:results}, we describe the quantum memory environment the PS agent is interacting with. The core component of the classical control (see Fig.~\ref{fig:agent-environment}) is the SQUAB algorithm~\cite{DelfosseNickerson2017} which simulates error correction on an arbitrary surface code. Note that error channels Eq.~(\ref{eq:errorZ})-~(\ref{eq:errorSpatial}) are simulated through erasure channels in SQUAB and hence yield error rates that differ slightly from their actual Pauli counterparts. Fortunately, the use of erasure channels is well motivated (see Sec.~\ref{sec:res_efficiency}).

%percept representation
The remainder of the environment is dedicated to the processing of percepts and actions. In the following, we give the details of what constitutes percepts and actions. Roughly, a percept represents the code structure of the current surface code. A surface code is fully described by the graph it is defined on. A graph can be represented by its adjacency matrix $A$ where rows represent vertices. Each vertex $v$ is represented by a set of edges adjacent to $v$, $A_v=(e_1,e_2,...)$. Edges are just labeled according to their first appearance in $A$. That is, $A_0=(0,1,...)$. Since we require a spatial resolution of the code, each vertex $v$ and plaquette $p$ is assigned a label $v,p\in\{0,1,2,...\}$. Edges, however, have no meaningful label. The specific code representation considered in this paper is hence a tuple of ordered adjacency matrices for the primal and dual graph.
%action representation
Actions have a straightforward representation that encodes Fig.~\ref{fig:mv}. Each action is a vector $a=(d, v, p_1, p_2)$. $d=0,1$ decides whether or not this action acts on the dual lattice. $v$ labels a vertex on the corresponding lattice and $p_1,p_2$ are two non-neighbouring plaquettes sharing edges with $v$ on the lattice's dual. In the spirit of our basic moves from Fig.~\ref{fig:mvb}, $a$ labels the vertex $v$ that is being split such that plaquettes $p_1,p_2$ become connected by an edge. For practical reasons, the actions are limited such that the connectivity of each vertex and plaquette can be at most 8 and at least 3. In that way, double edges are also excluded.

\section{Searching the space of surface codes}\label{app:search}

With this basic set of moves at our disposal (see Fig.~\ref{fig:mv}), one could be tempted to run an exhaustive search over the space of all surface codes. However, this problem is intractable, as we will see below.
Suppose we start from a topological code and search for a better code by increasing the code size. In order to preserve the topological structure of the code, additional qubits are injected by adding extra edges to the lattice or its dual as in Fig.~\ref{fig:mv}. Such actions increase the number of data qubits, while preserving the number of logical qubits. For instance, the $3 \times 3$ square lattice admits 18 primal actions and 18 dual actions (2 per square face).
A single extra qubit hence allows to reach 36 more topological codes. With two extra qubits, we already obtain 1440 new codes. Some of these codes may be identical but we treat them as distinct codes here in order to avoid the extra cost of comparing all new codes. That means that we navigate in a tree whose root is the initial code. Each node represents a topological code and the successors of a node are obtained by adding an edge as in Fig.~\ref{fig:mv}.

Imagine that we want to explore a region of the space of surface codes centered at the $3 \times 3$ square lattice. We use the software SQUAB~\cite{DelfosseIyerPoulin2016, SQUABweb} to perform 10,000 decoding trials for each code.  This costs roughly 0.018 seconds for a single code with 20 data qubits (using a processor 2.4 GHz Intel Core i5).
SQUAB returns an estimation of the maximum likelihood decoder performance over the quantum erasure channel for a given topological code (see Sec.~\ref{sec:res_efficiency}).
\begin{figure}[t!]
	\centering
	\includegraphics[width=.48\textwidth]{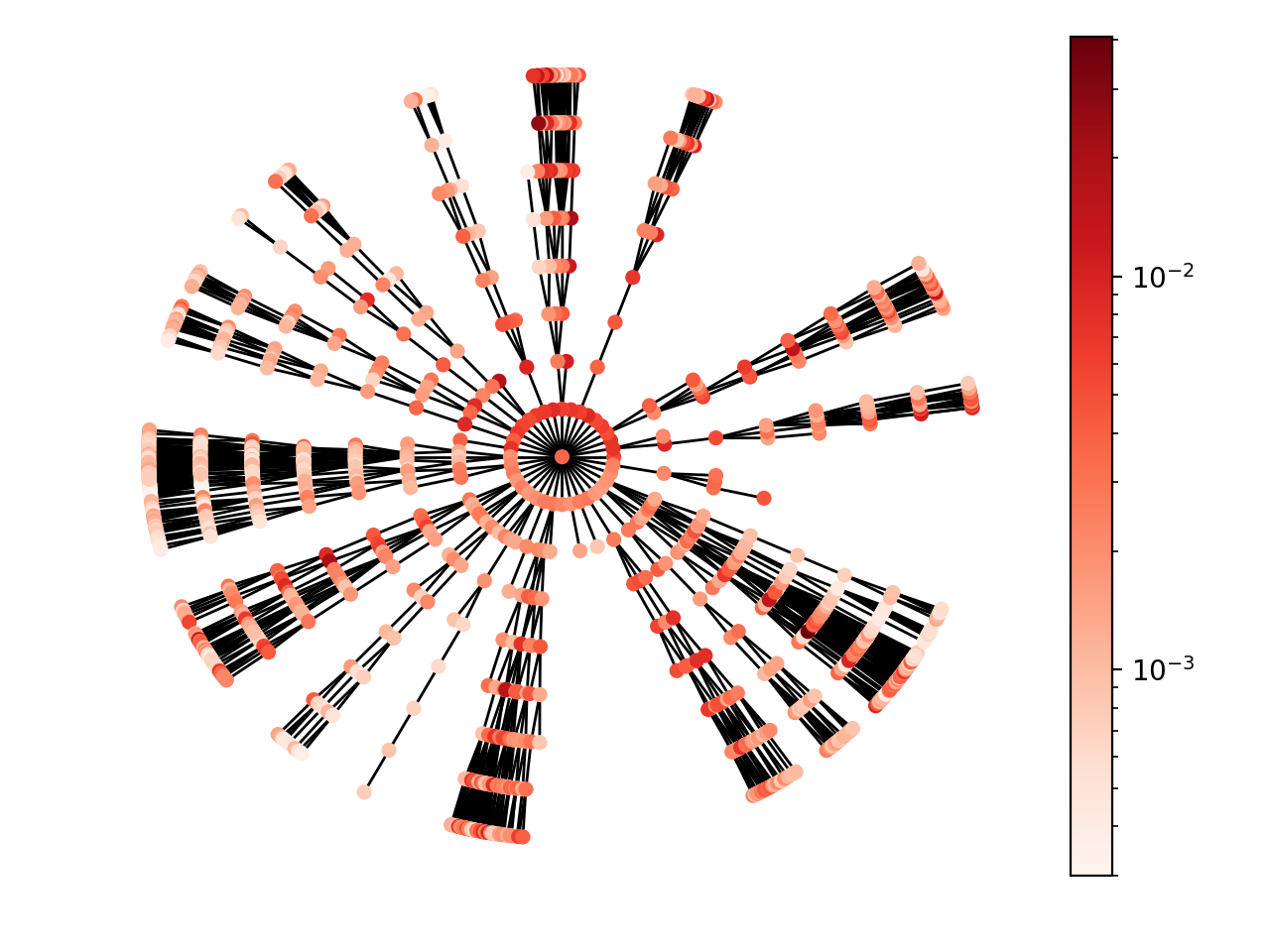}
	\caption{
		Exploration of random branches of the ball of radius $8$
		centered at the $3 \times 3$ surface code on a square lattice. Each node corresponds to a surface code on a torus. The color shading indicates the logical error rate.
		This rate is estimated using SQUAB with 10,000 trials per node at an error rate of $p_Z=0.15$ where $X$ errors are ignored.
		%The best result required $13$ additional qubits before the desired logical error rate $P_\mathrm{L}^{\mathrm{rew}}=0.001$ was reached.
	}
	\label{fig:branching_search}
\end{figure}
%%
%\vspace*{10mm}
%for a qubit erasure rate $p = 0.1$.
To simplify, we focus on the residual $Z$-error after correction and we ignore the $X$-part of the error. This is similar to the scenario considered in the main text with an error channel of the form Eq.~(\ref{eq:errorZ}).
Now, we want to explore all codes up to distance $r$ around the initial, $3 \times 3$ surface code on a square lattice (see Fig.~\ref{fig:sc}). In other words, we intend to explore the full ``ball'' of radius $r$ around the central node.
The number of codes $C(r)$ at distance $r$ from the root grows as follows,
	$C(0) = 1$,
	$C(1) = 36$,
	$C(2) = 1,440$,
	$C(3) = 62,893$,
	$C(4) = 2,961,504$.
Now, adding one extra qubit increases the number of codes by about a factor of 50.
Running SQUAB in a ball of radius 5 with 10,000 trials for each code would then require more than 30 days of computation. Similarly, going through all the codes obtained by adding up to 10 qubits would at least require 20 million years of CPU time.
This is without even counting the cost of generating the search graph and realizing all the moves.
The cost is even more discouraging when a more realistic noise model is considered increasing the simulation cost per node.
Hence, it is not reasonable to consider performing a numerical simulation of all the codes within the neighbourhood of an initial code in order to select the best candidate.

Exploring the whole neighbourhood of a code for a sufficiently large radius is hopelessly difficult. However, we can still use random search techniques to explore and visualize the environment, i.e. the space of surface codes. To this end, we explore a subset of branches and calculate the respective logical error rates at each node. We start by evaluating all the codes at distance one from the root to ensure a minimum number of nodes. Then we continue building each successor of a code in the search tree with probability $p_{\mathrm{expl}}$. Starting from a $3\times 3$ surface code, we are able to partially explore a ball of radius 8 with exploration probability $p_{\mathrm{expl}} = 0.03$ within a few minutes. This leads us to a set of $1,230$ randomly selected codes shown in Fig.~\ref{fig:branching_search}. We observe that moves which effectively increase the logical error rate are common. Moreover, after unnecessarily increasing the logical error rate, it is still possible to decrease the logical error rate again. In particular, it is very likely that the best codes are hidden among
unexplored branches of the search tree.
In order to explore in priority the most promising regions of the
space of all topological codes, we chose RL.
\begin{table}[!tb]\label{tbl:parameters}
	\setlength\tabcolsep{1.5pt} % let LaTeX compute intercolumn whitespace
	\footnotesize\centering
	\begin{tabular*}{\columnwidth}{|c|c|c|c|c|}
		\hline
		Figure	& $\eta$     & $\gamma$     & $\delta$    & SQUAB iterations  \\ \hline
		
		\ref{fig:resultiidZ}	& 0.05     & 0.01     &  0.01    &  1,000,000  \\ \hline
		\ref{fig:resultiidXZ}	&  0.01    & 0.01     &  0.01    &  1,000,000  \\ \hline
		\ref{fig:resultSpatial}	&  0.05    &  0.0006    & 0.001     & 1,000,000   \\ \hline
		\ref{fig:rewQuarter} trials $0$--$6,000$	&  0.05    &  0.01    & 0.01     & 1,000,000  \\ \hline
		\ref{fig:rewQuarter} trials $6,000$--$10,000$	&  0.05    &  0.0005    & 0.001     & 4,000,000   \\ \hline
		\ref{fig:comparison}	&  0.05    &  0.0006    &  0.001    &  1,000,000  \\ \hline
		\ref{fig:TL} trials $0$--$6,000$	&  0.05    &  0.01    & 0.01     & 1,000,000  \\ \hline
		\ref{fig:TL} trials $6,000$--$6,500$	&  0.05    &  0.0006    & 0.001     & 1,000,000   \\ \hline
	\end{tabular*}
	\caption{Parameters of the PS agent (see Appendix~\ref{app:PS}) and SQUAB algorithm as used for the various tasks considered in Sec.~\ref{sec:results} and~\ref{sec:on-off-line}. Two parameters remained the same for every setting: the softmax parameter $\beta=2$, and the immunity time $\tau=30$.}
\end{table}
%

%\newpage

\section{Parameters}\label{app:param}

For the results that are described in Sec.~\ref{sec:results} and displayed in Figs.~\ref{fig:resultiid}--\ref{fig:resultSpatial}, we deployed a PS agent as described in Sec.~\ref{sec:rl_PS} and Appendix~\ref{app:PS}. There are many parameters that can be tuned. For the specific results in this paper, the chosen parameters are displayed in Table.~\ref{tbl:parameters}. In particular, parameters can be tuned on-line (e.g. see Ref.~\cite{MakmalMelnikovDunjkoBriegel2016}). Considering a task as in Fig.~\ref{fig:rewQuarter} or~\ref{fig:TL}, it makes perfect sense to also change parameters between settings.

\bibliographystyle{apsrev4-1fixed_with_article_titles_full_names}
\bibliography{aqmbib}

\end{document}